%% file: main.tex
\documentclass[aps,prd,twocolumn,superscriptaddress,floatfix,longbibliography,preprintnumbers,nofootinbib]{revtex4-1} 
\usepackage{amsmath,amssymb} 
\usepackage[caption=false]{subfig}
\usepackage{scalerel}
\usepackage{float}
\DeclareUnicodeCharacter{202F}{FIXME}
\usepackage{lipsum}  
\usepackage{multirow}
\usepackage{bm} 
\usepackage{graphicx} 
\usepackage{comment} 
\usepackage{textcomp} 

\usepackage{booktabs}

\AtBeginDocument{%
  \heavyrulewidth=.08em
  \lightrulewidth=.05em
  \cmidrulewidth=.03em
  \belowrulesep=.65ex
  \belowbottomsep=0pt
  \aboverulesep=.4ex
  \abovetopsep=0pt
  \cmidrulekern=0.0em
  \defaultaddspace=.5em
}

\newcommand{\vect}[1]{\boldsymbol{\mathbf{#1}}}

\DeclareRobustCommand{\Sec}[1]{Sec.~\ref{sec:#1}}

\DeclareRobustCommand{\App}[1]{App.~\ref{app:#1}}

\DeclareRobustCommand{\Tab}[1]{Table~\ref{tab:#1}}

\DeclareRobustCommand{\Fig}[1]{Fig.~\ref{fig:#1}}

\DeclareRobustCommand{\Eq}[1]{Eq.~(\ref{eq:#1})}

\DeclareRobustCommand{\RRef}[1]{Ref.~\cite{#1}}
\DeclareRobustCommand{\RRefs}[1]{Refs.~\cite{#1}}

\usepackage{color}
\definecolor{darkgreen}{rgb}{0,0.7,0}

\usepackage{enumitem}

\usepackage{xcolor} 
\definecolor{lightgray}{gray}{0.6}
\definecolor{medgray}{gray}{0.4}
\definecolor{c1}{RGB}{249,65,68} 
\definecolor{c4}{RGB}{255,111,114} 
\definecolor{c2}{RGB}{0,168,50} 
\definecolor{c3}{RGB}{39,125,161} 
\definecolor{c5}{RGB}{157,111,255} 
\definecolor{c6}{RGB}{245,194,17} 

\newcommand{\neww}[1]{{{#1}}}

\usepackage{hyperref}

\usepackage{cleveref}
\hypersetup{
colorlinks=true,
urlcolor= blue,
citecolor=blue,
linkcolor= blue,
}

\newif\ifptitle
\newif\ifpnumber
\newcounter{para}
\newcommand\ptitle[1]{\par\refstepcounter{para}
{\ifpnumber{\noindent\textcolor{lightgray}{\textbf{\thepara}}\indent}\fi}
{\ifptitle{\textbf{[{#1}]}}\fi}}

\newcommand{\reft}[1]{Ref.~\cite{#1}}

\newcommand{\mytitle}{Comparing Point Cloud Strategies for Collider Event Classification}

\begin{document}

\title{\mytitle}
\author{Peter Onyisi}%
\email{ponyisi@utexas.edu}
\affiliation{%
  Department of Physics, University of Texas at Austin, Austin, TX 78712, USA
}%

\author{Delon Shen}

\email{delon@stanford.edu}

\thanks{Corresponding Author}
\affiliation{%
  Department of Physics, Stanford University, Stanford, CA 94305, USA
}%

\author{Jesse Thaler}
\email{jthaler@mit.edu}
\affiliation{
Center for Theoretical Physics, Massachusetts Institute of Technology, Cambridge, MA 02139, USA
}%
\affiliation{
The NSF AI Institute for Artificial Intelligence and Fundamental Interactions
}%

\preprint{MIT-CTP 5473}

\date{\today}

\begin{abstract}
In this paper, we compare several event classification architectures defined on the point cloud representation of collider events.
These approaches, which are based on the frameworks of deep sets and edge convolutions, circumvent many of the difficulties associated with traditional feature engineering.
To benchmark our architectures against more traditional event classification strategies, we perform a case study involving Higgs boson decays to tau leptons.
We find a 2.5 times increase in performance compared to a baseline ATLAS analysis with engineered features.
Our point cloud architectures can be viewed as simplified versions of graph neural networks, where each particle in the event corresponds to a graph node.
In our case study, we find the best balance of performance and computational cost for simple pairwise architectures, which are based on learned edge features.
\end{abstract}

\maketitle

\tableofcontents

\section{Introduction}

When analyzing data collected at the Large Hadron Collider (LHC), the ability to distinguish between specific production and decay channels is vital for picking out signal events among overwhelming backgrounds.
In the context of Higgs boson studies, the ATLAS and CMS collaborations rely heavily on dense neural networks (dNNs) \cite{CMS:2019lcn, CMS:2018hnq} and boosted decision trees (BDTs) \cite{ATLAS:2022yrq,ATLAS:2020evk,ATLAS:2018ynr,ATLAS:2017cen,ATLAS:2018kot,ATLAS:2017fak,ATLAS:2020fcp,ATLAS:2014vuz} for event classification.
These classifiers are typically trained on Monte Carlo (MC) simulated events to separate signal events from expected background processes.
Both dNNs and BDTs expect collider events to be represented by fixed-sized inputs.
Creating a robust fixed-sized representation of a collider event is challenging, however, often requiring hand engineering of a fixed number of features to distill relevant information from a variable number of particles.

\begin{figure*}
 \centering
 \def\svgwidth{\textwidth}
 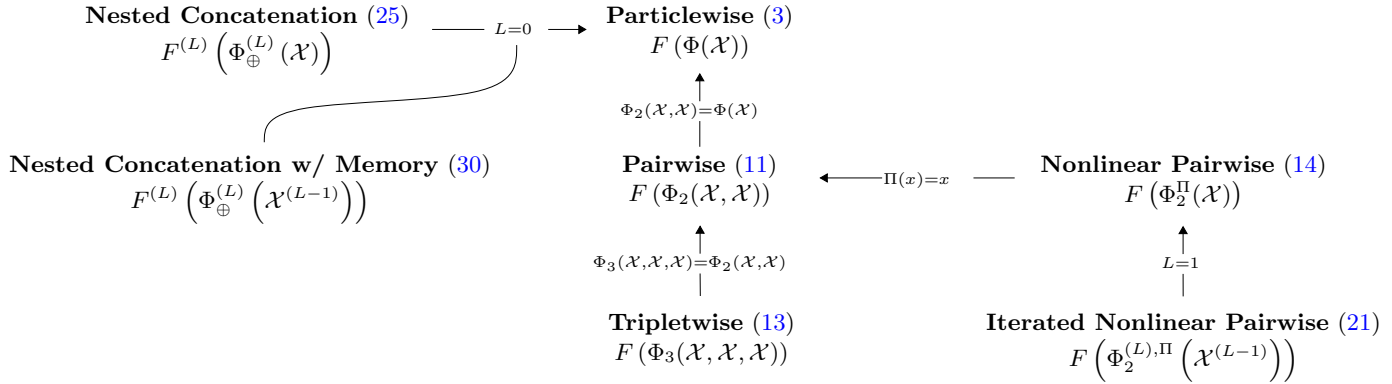
 \caption{
 Summary of the point cloud architectures studied in this paper, where the numbers refer to equations in the text.
 We also highlight limits where one architecture reduces to another.
 The Particlewise, Pairwise, and Tripletwise architectures are described in \Sec{local}.
 The Nonlinear Pairwise and Iterated Nonlinear Pairwise architectures are described in \Sec{iterated}.
The Nested Concatenation and Nested Concatenation w/ Memory architectures are described in \Sec{global} 
 }
 \label{fig:archs}
\end{figure*}

In this paper, we compare event classification architectures defined on \emph{point clouds}, which are a natural variable-sized representation of collider events.
Our architectures draw inspiration from two complementary approaches to point cloud processing.
The first is \emph{deep sets} \cite{Zaheer2017,Komiske:2018cqr}, which compute global information about an event based on permutation symmetric functions.
The second is \emph{edge convolutions} (EdgeConvs) \cite{Wang:2018nkf,Qu:2019gqs}, which compute local information associated with each particle and its neighbors.
We propose architectures which center around three different strategies for event classification:
\begin{itemize}
\item Emphasizing local information with \textit{multiple summations};
\item Improving latent representations of collider events from \textit{iterated convolutions}; and
\item Emphasizing global information through \textit{nested concatenation} of global features.
\end{itemize}
The nested concatenation structure provides an interesting alternative to the equivariant layers of \RRefs{Zaheer2017,Dolan:2020qkr}, with the same degree of expressivity.

To test our architectures, we perform a case study involving signal/background binary classification of Higgs boson decays to tau leptons.
This channel has been studied by ATLAS \cite{ATLAS:2018ynr, ATLAS:2020evk, ATLAS:2022yrq}, whose results we use as a baseline for comparison, and by CMS \cite{CMS:2014wdm, CMS:2017zyp, CMS:2021gxc, CMS:2022kdi}.
Based on our study, we recommend the \emph{pairwise architecture} in \Eq{pairwise_arch} below, which can be interpreted as either:
\begin{itemize}
\item A Deep Set acting on pairs of particles; or
\item A symmetric pooling over EdgeConvs.
\end{itemize}
Compared to more complex graph neural networks, this pairwise structure balances classification performance, computational efficiency, and conceptual simplicity.
We study the performance of the pairwise architecture as a function of the \textit{latent dimension} size, finding that even a single latent dimension outperforms the baseline ATLAS strategy.
Interestingly, we find that the discriminatory features identified by the pairwise architecture have some correlations with the traditionally selected hand-engineered features.

The point cloud representation and corresponding architectures have appeared before in the literature for various classification tasks \cite{Komiske:2018cqr, Qu:2019gqs, Moreno:2019bmu, Mikuni:2020wpr, Chakraborty:2020yfc, Shlomi:2020ufi, Shlomi:2020gdn, Dolan:2020qkr}. 
We refer to \RRefs{Duarte:2020ngm, Thais:2022iok} for a more thorough review of the use of point clouds in particle physics.
Using point clouds, collider events are represented as an unordered set of \(n\)-dimensional vectors, where each vector corresponds to a measured particle in that collision event.
We need a \textit{set} since there are a variable number of particles in each collision event.
This set is \textit{unordered} since there is no inherent ordering to the particles.%
\footnote{It is sometimes convenient to sort particles according to some measure of energy.  When comparing our point cloud architectures to fixed-sized input dNNs, we will sort over the particle transverse momenta ($p_T$).}
As discussed below, a key benefit of using architectures defined on point clouds is that it bypasses the traditional feature engineering game and associated combinatorial problems.

The remainder of this paper is organized as follows.
In \Sec{archs}, we describe our architectures and review their motivation and inspirations.
We make these architectures available along with example code on GitHub \cite{Shen2022}, and we describe the connection between nested concatenation and equivariant layers in \App{equiv_deep_set}.
In \Sec{exp}, we perform a classification case study of the $H\rightarrow\tau\tau$ decay channel, comparing our proposed architectures with a baseline ATLAS strategy, and we show visualizations of the latent space for the pairwise architecture.
Details of the neural network parameters are given in \App{params}.
We conclude in \Sec{conclusion} with a summary of our recommendations and areas for future exploration.

\section{Point Cloud Architectures}
\label{sec:archs}

In this section, we describe our point cloud architectures and their motivation.
We start by describing why the point cloud representation is more natural for event classification compared to traditional fixed-sized inputs.
We then review Deep Sets and EdgeConvs, which are the main inspirations for our architectures.
Following this, we describe our proposed architectures, as summarized in \Fig{archs}.

\subsection{Why Point Clouds for Event Classification}
\label{sec:why}

The point cloud representation of collider events avoids two of the key challenges when trying to construct robust fixed-sized inputs for event classification:
  \begin{itemize}
    \item \textbf{Combinatorial ambiguities}.
    One way of creating a fixed-sized representation of an event is to define high-level kinematic variables, but this can lead to combinatorial challenges.
    For example, if we find a good discriminatory feature that is derived assuming the final state has two $b$-tagged jets, but mistagging leads to three measured $b$-tagged jets, then one has to decide which of the three pairs should be used to compute the high-level feature.
    Similarly, if there is only one measured $b$-tagged jet, due to mistagging or kinematic acceptance, then the high-level feature is ill defined, even if in principle there is enough information available for event classification.
    \item \textbf{Truncation ambiguities}.
    Another way to create a fixed-sized representation of an event is to input the kinematics of a fixed number of particles.
    This, however, introduces a dependence on how the particles are ordered and which ones are truncated away.
    While some orderings, like taking the most energetic particles, have a physical motivation, they might not yield the best discrimination power, especially if particle correlations are relevant.
    There is also a question about how to properly pad events when there are fewer particles than the desired fixed-sized representation.
  \end{itemize}

The point cloud representation avoids these combinatorial and truncation ambiguities.
By enforcing permutation invariance, all possible particle combinations are automatically considered for event classification.
By allowing for variable-sized inputs, there is no need for ordering or truncation.
Of course, it is possible that cleverly engineered fixed-sized features could outperform generic point-cloud architectures, though this turns out not to hold for the case study in \Sec{exp}.
\neww{Graph neural networks are a popular approach to point cloud processing, but the simpler architectures studies in this paper also avoid these ambiguities with reduced computational complexity.}

The architectures below take as input a set of ${\color{c2}M}$ particles:
 \begin{equation}
 \mathcal X=\{x_1,\dots,x_{\color{c2}M}\}\subset \mathbb R^{\color{c1}n},
\end{equation}
  where each particle $x_i$ is described by ${\color{c1}n}$ features.
  These features could include particle characteristics like $p_T$ and $b$-tag score. 
  Let the classification of an event $\mathcal X$ be called:
  \begin{equation}
    f(\mathcal X) \subset \mathbb R^{{\color{c5}m}},
    \label{eq:general_notation}
  \end{equation} 
 where $f$ represents an event classification architecture for ${\color{c5}m}$ possible channels.
 For binary classification as studied in this paper, ${\color{c5}m}=2$.

\subsection{Review of Deep Sets}
  \label{sec:deepset}
  
Deep Sets are a way to parametrize permutation-symmetric functions with neural networks.
They were defined in \reft{Zaheer2017} and first introduced to the particle physics community in \reft{Komiske:2018cqr}, under the name of \emph{particle flow networks}.
Deep Sets have achieved state-of-the-art performance on various collider physics tasks, such as quark/gluon jet discrimination and boosted top tagging~\cite{Kasieczka:2019dbj}.

As shown in \reft{Zaheer2017}, a function $f(\mathcal X)$ operating on a set $\mathcal X$ is permutation invariant if (and, under certain assumptions, only if) it can be decomposed into the form:
    \begin{equation}
      f(\mathcal X) = F\left(\frac{1}{\color{c2} M} \sum_{i=1}^{{\color{c2} M}} \Phi(x_i) \right) .
    \label{eq:inv_deep_set}
    \end{equation}
Each particle $x_i\in \mathcal X$ is transformed into a latent representation of dimension ${\color{c3}\ell}$ by a function:
\begin{equation}
\Phi:\mathbb R^{\color{c1}n}\rightarrow \mathbb R^{\color{c3}\ell}.
\end{equation}
The particle-wise outputs $\Phi(x)$ are averaged over and processed by an event-wise function:
\begin{equation}
F:\mathbb R^{\color{c3}\ell} \rightarrow \mathbb R^{\color{c5}m}.
\end{equation}
To approximate the optimal event classifier, the functions $\Phi$ and $F$ are parametrized by neural networks.

The factor of $1/{\color{c2} M}$ in \Eq{inv_deep_set} differs from the presentation in \RRefs{Zaheer2017,Komiske:2018cqr}, where sum pooling (instead of average pooling) was the default.
Average pooling simplifies some of the later notation, and we use this pooling operation to define the action of $\Phi$ on the set $\mathcal X$:
\begin{equation}
\label{eq:phi_on_set}
\Phi(\mathcal{X}) \equiv \frac{1}{\color{c2} M} \sum_{i=1}^{{\color{c2} M}} \Phi(x_i).
\end{equation}
This notation makes more clear that the function $\Phi$ effectively maps the entire set $\mathcal X$ into a latent representation of dimension ${\color{c3}\ell}$.

\subsection{Review of Edge Convolutions}
  \label{sec:edgeconv}
  EdgeConvs are a way to incorporate local neighborhood information within point clouds for learning tasks.
  They were introduced in \reft{Wang:2018nkf} and first used in particle physics by \reft{Qu:2019gqs}, under the name of ParticleNet.
  Architectures incorporating EdgeConvs have also achieved state-of-the-art performance for collider tasks~\cite{Kasieczka:2019dbj}.

  The EdgeConv mechanism in \reft{Wang:2018nkf} boils down to the following transformation of a particle \(x_i\in\mathcal X\):
      \begin{equation}
        \label{eq:latent2}
        \Phi_2(x_i,\mathcal X) \equiv  \frac{1}{{\color{c2} M}}\sum_{j=1}^{{\color{c2} M}} \Phi_2(x_i,x_j),
      \end{equation}
      where the notation mirrors that of \Eq{phi_on_set}.
      Here, the particle $x_i$ is transformed into a latent representation through a function $\Phi_2$ based on pairwise information: 
      \begin{equation}
      \Phi_2(x_i,x_j):\mathbb R^{\color{c1}n}\times \mathbb R^{\color{c1}n}\rightarrow\mathbb R^{\color{c3}\ell}.
      \label{eq:latent2dim}
      \end{equation}
      Like before, ${\color{c3}\ell}$ is the latent dimension, and  $\Phi_2$ is parametrized by a neural network.
      For later purposes, we can define $\Phi_2$ acting on two sets via:
      \begin{align}
      \nonumber
      \Phi_2(\mathcal X,\mathcal X) & \equiv  \frac{1}{{\color{c2} M}}\sum_{i=1}^{{\color{c2} M}} \Phi_2(x_i,\mathcal X) \\ & \equiv  \frac{1}{{\color{c2} M}^2}\sum_{i=1}^{{\color{c2} M}} \sum_{j=1}^{{\color{c2} M}} \Phi_2(x_i,x_j).\label{eq:phi2_on_sets}
      \end{align}

      The EdgeConv operation can be iterated and combined with various nonlinearities  and pooling operations to create a \textit{Dynamic Graph Convolutional Neural Network} (DGCNN). These were used for various computer vision and graphics tasks in \reft{Wang:2018nkf} and shown to achieve strong performance on several standard benchmark tasks.  Furthermore during these benchmarks, it was shown that DGCNN's achieved the best trade off between the number of parameters and the runtime of the model.

\subsection{Multiple Summation}
          \label{sec:local}

          We now describe our first set of proposed architectures, based on simple summations.
          The \textbf{Particlewise architecture} is a direct application of the Deep Set formalism in \Eq{inv_deep_set}:
          \begin{align}
	\nonumber f(\mathcal X) &= F\left(\frac{1}{\color{c2} M} \sum_{i=1}^{{\color{c2} M}} \Phi(x_i) \right) \\
            &= F\Big( \Phi(\mathcal X) \Big),
            \label{eq:particlewise_arch}
          \end{align}
          where in the last line, we are using the notation from \Eq{phi_on_set}.
          The set of particles $\mathcal X$ are transformed into a latent \({\color{c3}\ell}\)-dimensional representation by $\Phi$, which is further processed into the output by $F$.
          Because the particle-wise function $\Phi$ can only see one particle at a time, this architecture is inefficient at capturing local information in the vicinity of each particle.

          We can improve our ability to process local information using the \textbf{Pairwise architecture}, which acts on all \textit{pairs} of particles in an event:%
          \footnote{We thank Patrick Komiske and Eric Metodiev for foundational discussions related to this architecture.}
          \begin{align}
        \nonumber
	f(\mathcal X) & = F\left(\frac{1}{{\color{c2} M}^2} \sum_{i=1}^{{\color{c2} M}}\sum_{j=1}^{{\color{c2} M}}\Phi_2(x_i,x_j) \right)\\
      \nonumber &= F\left( \frac{1}{{\color{c2}M}}\sum_{i=1}^{{\color{c2}M}} \Phi_2(x_i,\mathcal X) \right) \\
        &= F \Big( \Phi_2(\mathcal X,\mathcal X)\Big).
        \label{eq:pairwise_arch} 
      \end{align}
          Using \Eq{phi2_on_sets}, the three different notations above correspond to three different ways of thinking about this architecture.
          The first line corresponds to applying the Deep Sets formalism in \Eq{inv_deep_set} to the set of (ordered) particle pairs.%
          \footnote{With average pooling, we could choose to work with unordered pairs and make $\Phi_2(x_i,x_j)$ symmetric in its arguments, but using ordered pairs often yields simpler manipulations.}
          The second line corresponds to applying the EdgeConvs in \Eq{latent2} to each particle, and then performing average pooling and postprocessing.
          The last line emphasizes that the role of $\Phi_2$ is to map particle pairs to an \({\color{c3}\ell}\)-dimensional latent space, which makes this local pairwise information directly accessible when constructing a latent representation.
          If we choose $\Phi_2(x_i,x_j)=\Phi(x_i)$, such that the $x_j$ input is ignored, then the Pairwise architecture reduces to the Particlewise one.

         The natural generalization of the above constructions is the \textbf{Tripletwise architecture}, which involves a function that maps triplets of particles into a latent space:
          \begin{equation}
       \Phi_3(x_i,x_j,x_k):\mathbb R^{\color{c1}n}\times \mathbb R^{\color{c1}n} \times \mathbb R^{\color{c1}n} \rightarrow\mathbb R^{\color{c3}\ell}.
      \end{equation}
         Mirroring the notation in  \Eq{phi2_on_sets}, we can define this architecture in multiple ways:
         \begin{align}
        \nonumber
	f(\mathcal X) &= F \left( \frac{1}{{\color{c2}M}^3}\sum_{i=1}^{{\color{c2}M}}\sum_{j=1}^{{\color{c2}M}}\sum_{k=1}^{{\color{c2}M}} \Phi_3(x_i,x_j,x_k) \right)\\
        \nonumber &= F \left( \frac{1}{{\color{c2}M}^2}\sum_{i=1}^{{\color{c2}M}}\sum_{j=1}^{{\color{c2}M}} \Phi_3(x_i,x_j,\mathcal X) \right)\\
        \nonumber &= F \left( \frac{1}{{\color{c2}M}}\sum_{i=1}^{{\color{c2}M}}\Phi_3(x_i,\mathcal X,\mathcal X)\right)\\
        &= F \Big(\Phi_3(\mathcal X,\mathcal X,\mathcal X)\Big).
        \label{eq:triplesum}
      \end{align}
         This tripletwise structure makes available even more local information when constructing a latent representation of an event.
         If we choose $\Phi_3(x_i,x_j,x_k)=\Phi_2(x_i,x_j)$, such that the $x_k$ input is ignored, then the Tripletwise architecture reduces to the Pairwise one.
         We found it impractical to use architectures with more nested summations due to their heavy computational cost.

      \subsection{Iterated Convolutions}
      \label{sec:iterated}
      
      One way to make neural networks more expressive is to introduce more nonlinearities.
      This is the motivation for our iterated convolution architectures, which can be viewed as a special case of DGCNNs~\cite{Wang:2018nkf}.

      A natural evolution of the Pairwise architecture in \Eq{pairwise_arch} involves inserting an additional non-linear function $\Pi$ implemented with a neural network between the two sums, or equivalently after the EdgeConv layer, which  yields the \textbf{Nonlinear Pairwise architecture}:
       \begin{align}
        \nonumber
        f(\mathcal X) & = F \left( \frac{1}{{\color{c2}M}}\sum_{i=1}^{{\color{c2} M}} \Pi\left( \frac{1}{{\color{c2} M}}\sum_{j=1}^{{\color{c2} M}}\Phi_2(x_i, x_j) \right)  \right)\\
        \nonumber & = F \left( \frac{1}{{\color{c2}M}}\sum_{i=1}^{{\color{c2} M}} \Pi \big( \Phi_2(x_i,\mathcal X)\big) \right) \\
         &= F \Big(\Phi_2^\Pi(\mathcal X)  \Big).
        \label{eq:edgeconvnl}
      \end{align}
      Here, we have introduced the notation
      \begin{equation}
          \Phi_2^\Pi(\mathcal X) \equiv \frac{1}{{\color{c2}M}}\sum_{i=1}^{{\color{c2}M}} \Pi \big( \Phi_2(x_i,\mathcal X) \big) ,
          \label{eq:phipi}
        \end{equation}
      which emphasizes that we are still transforming the point cloud into a latent representation of dimension ${\color{c3}\ell}$.
      Note that the output of $\Phi_2$ need not be ${{\color{c3}\ell}}$ dimensional as in \Eq{latent2dim}, since now the latent representation is constructed through both \(\Pi\) and \(\Phi_2\):
     \begin{align}
       \Phi_2(x_i, x_j)&:  \mathbb R^{{\color{c1}n}}\times \mathbb R^{{\color{c1}n}} \to \mathbb R^{{\color{c3}\ell'}}, \\
       \Pi &:  \mathbb R^{{\color{c3}\ell'}} \to \mathbb R^{{\color{c3}\ell}},
      \end{align}
      where ${{\color{c3}\ell'}}$ and ${{\color{c3}\ell}}$ could differ.

      We can increase the expressivity of the latent representations by iteratively applying nonlinearities and EdgeConvs.
      Let
       \begin{equation}
      \mathcal X^{(0)} \equiv \mathcal X
      \end{equation}
      be the original point cloud.
      We then define the point cloud at depth $d$ as
      \begin{equation}
        \mathcal X^{(d)}  = \left\{ x_1^{(d)},\dots, x_{{\color{c2}M}}^{(d)} \right\},
      \end{equation}
      where the set size ${\color{c2}M}$ matches the original point cloud.
      The $i$-th element at depth $d$ is the result of applying a nonlinearity $\Pi^{(d)}$ to an EdgeConv defined by $\Phi_2^{(d)}$:
      \begin{equation}
      \label{eq:nestedeq}
          x^{(d)}_i =  \Pi^{(d)}\left( \Phi_2^{(d)}\left(x_i^{(d-1)},\mathcal X^{(d-1)}  \right)  \right).
      \end{equation}
      The \textbf{Iterated Nonlinear Pairwise architecture} at total depth $L$ is achieved by average pooling over these elements and then postprocessing: 
      \begin{align}
        \nonumber
        f(\mathcal X) &= F \left(\frac{1}{{\color{c2}M}} \sum_{i=1}^{{\color{c2} M}} x^{(L)}_i \right) \\
        \nonumber &= F \left( \frac{1}{{\color{c2}M}}\sum_{i=1}^{{\color{c2}M}}\Pi^{(L)} \left( \Phi_2^{(L)} \left( x_i^{(L-1)}, \mathcal X^{(L-1)} \right)  \right)  \right) \\
                               &\equiv F \left( \Phi_2^{(L),\Pi}(\mathcal X^{(L-1)}) \right).
    \label{eq:iteratedecnl}
    \end{align}
      In the last line, we are using the same notation as \Eq{phipi}, to emphasize that $\Phi_2^{(L),\Pi}$ transforms the point cloud into an ${{\color{c3}\ell}}$-dimensional latent representation. 
      When \(L=1\), this reduces to the Nonlinear Pairwise architecture from \Eq{edgeconvnl}.
      A version of this architecture is also possible for the tripletwise case, but we shall not pursue it due to its heavy computational cost.

\subsection{Nested Concatenation of Global Features}
      \label{sec:global}
      
 Our final class of architectures combines global information about the whole event with local information about particles.
 Let $\hat f(\mathcal X)$ be a permutation invariant function, which captures global information about the collider event $\mathcal X$.
 For example, $\hat f(\mathcal X)$ could simply be a Deep Set applied to the set of particles:
      \begin{equation}
        \hat f(\mathcal X) = \hat{F}\left( \frac{1}{{\color{c2}M}}\sum_{i=1}^{{\color{c2} M}}\hat{\Phi}(x_i) \right).
      \end{equation}
The global information from $\hat f(\mathcal X)$, can then be concatenated $(\oplus$) with local features associated with each particle:
      \begin{equation}
        f(\mathcal X) = F \left( \frac{1}{{\color{c2}M}}\sum_{i=1}^{{\color{c2} M}} \Phi\big(x_i\oplus \hat f(\mathcal X) \big) \right).
      \label{eq:pool}
      \end{equation}
This concatenation is a conceptually simple way to let individual particles see global information about the point cloud.

The \textbf{Nested Concatenation architecture} iterates the structure in \Eq{pool} for $L$ times.
Let the base case be a standard Deep Set:
      \begin{equation}
        f^{(0)}(\mathcal X) = F^{(0)} \left(\frac{1}{{\color{c2}M}} \sum_{i=1}^{{\color{c2}M}} \Phi^{(0)}(x_i) \right).
        \label{eq:nested_base}
      \end{equation}
At level $d$, we have:
        \begin{align}
          \nonumber
          f^{(d)}(\mathcal X)&= F^{(d)} \left( \frac{1}{{\color{c2}M}}\sum_{i=1}^{{\color{c2} M}} \Phi^{(d)} \left( x_i \oplus f^{(d-1)}(\mathcal X) \right)  \right) \\
                             &= F^{(d)} \left( \Phi^{(d)}_\oplus(\mathcal X) \right).
        \label{eq:nested}
        \end{align}
        In the last line, we have introduced the notation 
        \begin{align}
          \Phi_\oplus^{(d)} \left(\mathcal X  \right) &= \frac{1}{{\color{c2}M}}\sum_{i=1}^{{\color{c2}M}} \Phi^{(d)} \left( x_i \oplus f^{(d-1)}\left( \mathcal X \right)  \right),
          \label{eq:phioplus}
        \end{align}
        which emphasizes that $\Phi_\oplus^{(d)}$ transforms the point cloud into a latent representation of dimension ${{\color{c3}\ell}}$, just as in the previous architectures.
        The final architecture at total level $L$ is
      \begin{equation}
      f(\mathcal X) \equiv f^{(L)}(\mathcal X),
      \end{equation}
such that \Eq{nested} reduces to \Eq{inv_deep_set} when $L = 0$.

To allow for more dynamic manipulation of the point cloud information, we define the \textbf{Nested Concatenation with Memory architecture}.
In this architecture, the intermediate latent representations of particles generated by $(d-1)$-th nested layer are also utilized in the $d$-th nested layer.
We define the same base case function as \Eq{nested_base}, with the base case set as $\mathcal X^{(0)} \equiv \mathcal X$.
After $d$ nested layers, the set takes the form
      \begin{equation}
        \label{eq:gen_nested_xd}
        \mathcal X^{(d)} = \left\{ x_1^{(d)},\dots,x_{{\color{c2}M}}^{(d)} \right\},
      \end{equation}
where the $i$-th element is determined via:
      \begin{align}
      \label{eq:concat_for_AppA}
        x^{(d)}_i = \Phi^{(d)}\left( x_i^{(d-1)}\oplus f^{(d-1)}\big(\mathcal X^{(d-1)}\big) \right).
      \end{align}
   The function $f^{(d)}(\mathcal X^{(d)})$ sums and processes the latent representation $\mathcal X^{(d)}$ with a function \(F^{(d)}\):
      \begin{align}
        \nonumber
        f^{(d)}(\mathcal X^{(d)}) &= F^{(d)} \left( \frac{1}{{\color{c2}M}}\sum_{i=1}^{{\color{c2}M}}x_i^{(d)} \right) \\
                                  &= F^{(d)} \left( \Phi_{\oplus}^{(d)}\big( \mathcal X^{(d-1)} \big)  \right) .
        \label{eq:gen_nested}
      \end{align}
The function $\Phi_{\oplus}^{(d)}$ is the same as in \Eq{phioplus} but now applied to the set $\mathcal X^{(d-1)}$.
Like before, we let
      \begin{equation}
         f(\mathcal X) \equiv f^{(L)}(\mathcal X^{(L)})
      \end{equation}
      be the classification of collider event $\mathcal X$ at level $L$.

It is worth mentioning that \RRef{Zaheer2017} defined an alternative method to incorporate global information based on permutation equivariance; this structure was used for jet tagging in \RRef{Dolan:2020qkr}.
In \App{equiv_deep_set}, we show that permutation equivariant Deep Sets are a special case of our Nested Concatenation architecture.
We prefer to use the more general concatenation structure, though, due to its flexibility.

\section{Event Classification Case Study}
\label{sec:exp}

In this section, we describe the setup and results of our event classification case study to benchmark our proposed architectures against more traditional methods.
We start with a description of the signal and background processes that will be the context for our case study.
We then describe how we generate synthetic data sets and preprocess the inputs for both traditional architectures and our proposed architectures.
Following this, we present several performance metrics of the tested architectures and advocate for the Pairwise architecture as the best balance between computational cost and performance.
We then perform a latent dimension study of the Pairwise architecture and visualize the separation of signal and background events in the latent space. 
Finally, we examine correlations between the features found to be useful to represent collider events by our Pairwise architecture and the hand-engineered features chosen by the ATLAS collaboration.

\subsection{Signal and Backgrounds Processes}
\label{sec:physical}

\begin{figure*}
  \subfloat[\label{fig:ttH}]{
    \includegraphics[height=0.4\textheight]{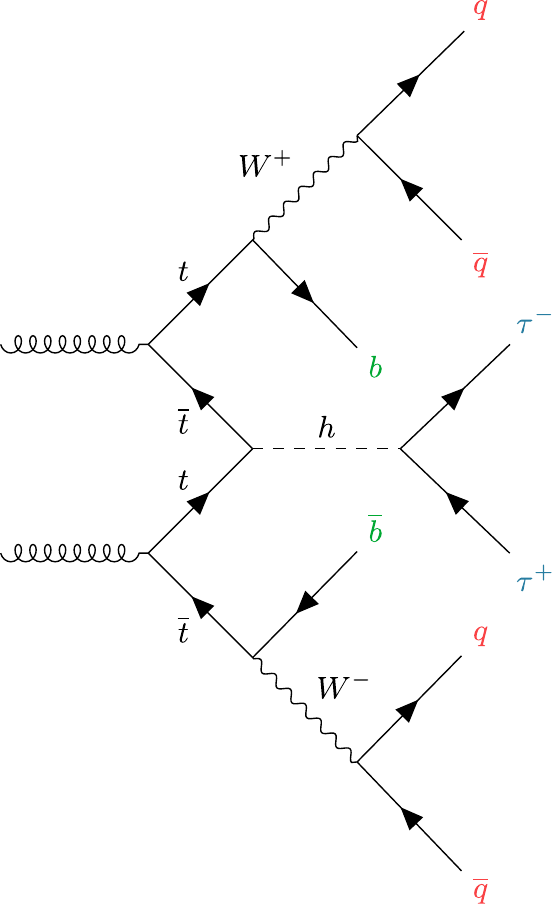}
    
  }
  \hspace{6em}
  \subfloat[\label{fig:ttbar}]{
    \includegraphics[height=0.4\textheight]{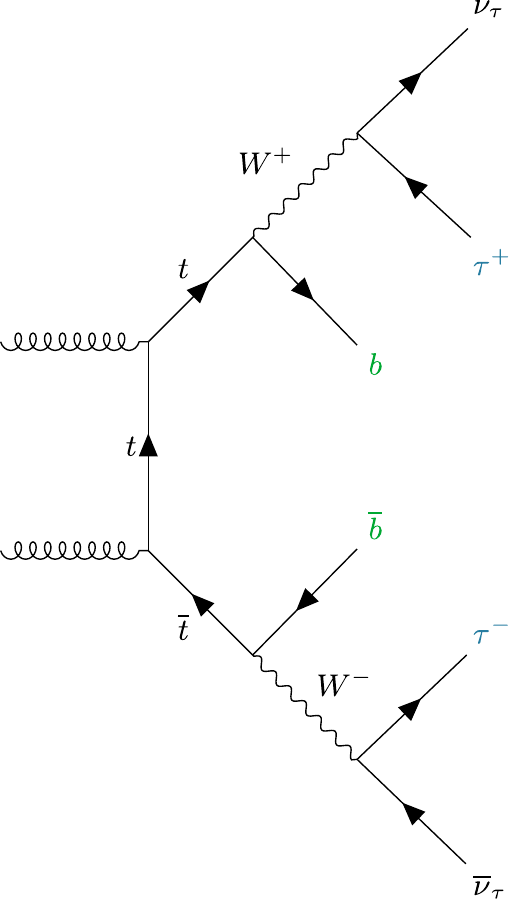}
  }
  \caption{
  A schematic of the (a) signal and (b) background process considered in our event classification case study.
  The signal process in (a) involves the production of a Higgs boson decaying to \(\tau\tau\) associated with the production of a pair of top quarks where both top quarks and both \(\tau\) leptons decay hadronically.
  The background process in (b) involves the production of a pair of top quarks where both tops decay to \(\tau\nu b\) and both \(\tau\)s decay hadronically.
  The background process can mimic the signal process if there are additional jets from initial-state radiation, which was found to be a significant effect in a recent ATLAS analysis of \(H\rightarrow\tau\tau\) \cite{ATLAS:2022yrq}.
  }
  \label{fig:physical}
\end{figure*}

\ptitle{Description of event classification experiment + event generation}

Our case study is based on a problem relevant to analyzing the $H\rightarrow \tau\tau$ decay channel \cite{ATLAS:2018ynr, ATLAS:2020evk, ATLAS:2022yrq}.
For leptonic Higgs boson decays, the $H\rightarrow\tau^+\tau^-$ channel has the largest branching ratio of 6.3\% \cite{Djouadi:1997yw,LHCHiggsCrossSectionWorkingGroup:2016ypw} which makes this channel a prime candidate to study the Yukawa-Higgs mechanism for mass generation.
The presence of neutrinos in the final state of this process, however, degrades the resolution of the measured Higgs boson four momentum.
\neww{This degraded resolution makes the signal process much more difficult to distinguish from background processes, thereby motivating a machine learning approach.}

To tackle the challenge of identifying the $H\rightarrow\tau^+\tau^-$ final state, one typically isolates different event topologies and studies them individually. 
One $H\rightarrow\tau\tau$ topology of interest---which will serve as the signal process in our classification case study---is the production of a Higgs boson associated with a pair of top quarks where both top quarks and both $\tau$ leptons decay hadronically.
We denote this signal process as 
\begin{equation}
  t\overline{t}(H\rightarrow\tau\tau)\textrm{ or }t\overline{t}H \textrm{ for short.}
\end{equation}
A schematic of this process is shown in \Fig{ttH}.
Ideally, the final state for this process would result in two {\color{c3}$\tau$-tagged jets}, two {\color{c2}$b$-tagged jets}, and four {\color{c1}additional jets from \(W^\pm\) decay}.
This channel was considered by ATLAS in \RRef{ATLAS:2022yrq} and by CMS in \RRef{CMS:2020mpn}.

The main background process that mimics this $t\overline{t}H$ signature---and significantly hinders the analysis of this channel \cite{ATLAS:2022yrq}---is the production of a top-antitop quark pair where each top quark decays as $t\rightarrow\tau\nu b$ and both $\tau$ decay hadronically.
We denote this background process as 
\begin{equation}
  t\overline{t}(\rightarrow\tau\nu b)  \textrm{ or }t\overline{t}\textrm{ for short.}
\end{equation}
A schematic of this process is shown in \Fig{ttbar}.
The $t\overline{t}$ channel can mimic the signature of the ideal $t\overline{t}H$ process if there are four additional jets from gluons radiated before the hard scattering.
In our case study, we focus on distinguishing between $t\overline{t}H$ events and $t\overline{t}$ events.
A full analysis, of course, would consider multiple Higgs production topologies.

\neww{These two channels both exhibit high multiplicity final states with a diverse range of final state objects. This leads to many potential combinatorial reconstructions and a high probability of detecting extra or losing relevant objects. These characteristics make manually constructing features difficult and motivates the need for flexibility in the number of input objects. Our case study is therefore representative of situations where we hope to make gains from using point-cloud-based architectures, which naturally account for combinatorial ambiguities and incomplete reconstruction.  The particular processes we study are among the highest multiplicity channels currently analyzed at the LHC.}

\subsection{Data Generation}
\label{sec:data}

Following the $t\overline{t}(H\rightarrow\tau\tau)$ analysis strategy in \RRef{ATLAS:2022yrq}, we select events that satisfy the following properties:
\begin{itemize}
  \item Two visible $\tau$-tagged jets, with kinematic conditions: 
   \begin{itemize}
    \item \({\rm max}(\{p_T^\tau\})>40\) GeV,
    \item \({\rm min}(\{p_T^\tau\})>30\) GeV,
    \item \(0.6<\Delta R_{\tau\tau}<2.5\),
    \item \(|\eta|\leq 2.5\) 
    \item \(|\Delta \eta_{\tau\tau}|<1.5\),
    \item \(0.1<x_1,x_2<1.4\) (defined below);
  \end{itemize}
  \item ($\geq$ 5 jets and $\geq$ 2 b-tags) \textbf{or} ($\geq$ 6 jets and $\geq$ 1 b-tags), with kinematic conditions for 
    \begin{itemize}
      \item the leading (non-\(\tau\)) jet:
  \begin{itemize}
    \item \(p_T>70\) GeV,
    \item \(|\eta|<3.2\); and
  \end{itemize}
\item other jets: 
  \begin{itemize}
    \item \(p_T \geq 20\) GeV,
    \item \(|\eta|\leq 5\); 
  \end{itemize}
    \end{itemize}  \item $\leq 15$ jets total.\footnote{This condition was not present in the original analysis of \RRef{ATLAS:2022yrq}.  The analysis team reported, however, than there no events in the data set that contained more than 15 jets.}
\end{itemize}
Here, the transverse momenta (\(p_T\)) and pseudorapidities \((\eta)\) are defined with respect to the beamline, $\Delta R^2 = \Delta \eta^2 + \Delta \phi^2$ is the distance between objects in the pseudorapidity-azimuth ($\eta$-$\phi$) plane, and $x_1$, $x_2$ are the momentum fractions carried away by visible \(\tau\) decay products as computed by the collinear approximation~\cite{Ellis:1987xu, Elagin:2010aw, Konar:2016wbh}.

For both the signal and background channels, we generate events with \textsc{MadGraph 5 v3.1.1}~\cite{Alwall:2011uj} and \textsc{Pythia 8.245}~\cite{Sjostrand:2014zea}.
These events are passed through the \textsc{Delphes 3.5.0}~\cite{deFavereau:2013fsa} detector simulation with the ATLAS card.%
\footnote{Following the working point of \RRef{ATLAS:2022yrq}, this ATLAS card is modified so that $\tau$-tagging is at 100\% efficiency.  This is because we apply a two $\tau$ preselection, so there is no reason to simulate tau finding inefficiency.}
Jets are then clustered with the $R=0.4$ anti-$k_T$ algorithm \cite{Cacciari:2008gp} using the \textsc{FastJet 3.3.4} \cite{Cacciari:2011ma} package.
From all of the generated events, we extract 80k events from each channel that satisfy the event selection criteria, such that we have balanced data sets.
For the machine learning study, we split each data set into \(70\%\) for training and \(30\%\) for testing.

\subsection{Data Processing}
\label{sec:process}

The ATLAS analysis in \RRef{ATLAS:2022yrq} is based on the following engineered features:
\begin{enumerate}
  \item \(\Sigma p_T^{\rm jets}\): Scalar sum of all jets $p_T$;
  \item \(M_{\hat{W}}\): Invariant mass of the dijet with invariant mass closest to $W$ boson mass;
  \item \(\Delta R_{\rm min}\): Smallest $\Delta R$ between any two jets;
  \item \(M_{\hat{t}}\): Invariant mass of the trijet with invariant mass closest to top quark mass;
  \item \(\Delta R_{\tau\tau}\): $\Delta R$ between two $\tau$-tagged jets;
  \item \(|\Delta \eta_{\tau\tau}|\): $|\Delta \eta|$ between two $\tau$-tagged jets;
  \item \(p_T^{\tau\tau}\): The $p_T$ of the $\tau\tau$ dijet; and
  \item \(E_T^{\rm miss}\): Missing transverse energy $E_T^{\rm miss}$.
\end{enumerate}
These ATLAS features are used as inputs for a BDT to mimic the ATLAS analysis.
We also use these ATLAS features to train a dNN, which yields comparable performance to the BDT.
For discrimination between \(t\overline{t}H\) and other background processes such as \(Z+{\rm jets}\), alternative features are chosen by ATLAS.

For the point cloud architectures, we perform the following preprocessing of the inputs.
Let $K_{\rm tot}$ be the scalar sum of the kinematic quantity $K$ for all particles in an event and $\tilde K_{\rm tot}$ be the kinematic quantity $K$ derived from the sum of the four-momenta of all particles in the event.
We also define $K_i$ as the kinematic feature for object $i$ in the event.
Each event is represented as a set of particles with the following kinematic features representing each particle:  
\begin{equation}
    {\rm Event}=\left\{ {\rm particle}_i \right\} \equiv \left\{   \left[\begin{matrix}  \log\big( {E_{i}} / {\tilde E_{\rm tot} } \big) \\  \log\left( {M_{i}}/{M_{\rm tot} } \right)\\ \eta_{i} \\  \phi_{i}\\  \log\left( {p^{i}_T}/{p^{\rm tot}_T} \right) \\  \left( b_{\rm tag} \right) _i\\  \left( \tau_{\rm tag} \right)_i   
  \end{matrix}   \right]\right\}.
  \label{eq:rep}
\end{equation}
Here, $E$ is energy in the lab frame, $M$ is invariant mass, and
\begin{align*}
    b_{\rm tag}&: \textrm{ \(1\) if particle is \(b\)-tagged and \(0\) if not},\\
    \tau_{\rm tag}&: \textrm{ \(1\) if a particle is \(\tau\)-tagged and \(0\) if not}.
\end{align*}
We found that taking the logarithm of the dimensionful features improved the training across architectures.

One additional hand-engineered feature we must consider is the  di-tau invariant mass $M_{\tau\tau}$.
This feature is left out of standard ATLAS training but discoverable by the point cloud architectures. 
Because the di-tau mass would peak at the Higgs mass for signal events but not for background events, it is expected to be a good discriminant.  
In the context of the ATLAS analysis, this feature is intentionally left out so that it can be used as a sanity check on the classification.
To get a more complete performance comparison for our case study, we concatenate $M_{\tau\tau}$ with the previously described ATLAS features to be used as input to a BDT.
Specifically, we include:
\begin{enumerate}
\setcounter{enumi}{8}
\item $M_{\tau\tau}^{\rm Coll}$:  The reconstructed di-tau invariant mass using the collinear approximation~\cite{Ellis:1987xu, Elagin:2010aw, Konar:2016wbh} to account for the energy carried off by neutrinos.
\end{enumerate}
The ATLAS analysis in \RRef{ATLAS:2022yrq} uses the Missing Mass Calculator~\cite{Elagin:2010aw}, but we chose to test against the collinear approximation instead for its relative simplicity.

As a cross check, we train a dNN that takes as input a $p_T$-sorted and flattened version of the event representation in \Eq{rep} with padding to make each event have 15 objects.
We call this a \textit{flattened point cloud}. 
By training a dNN on the flattened point cloud, we can assess whether the improvement in performance we find from the point cloud architectures is truly due to how we structure the architectures rather than just because of an increase in available information.

\begin{figure}[t]
 \begin{center}
   \includegraphics[width=\linewidth]{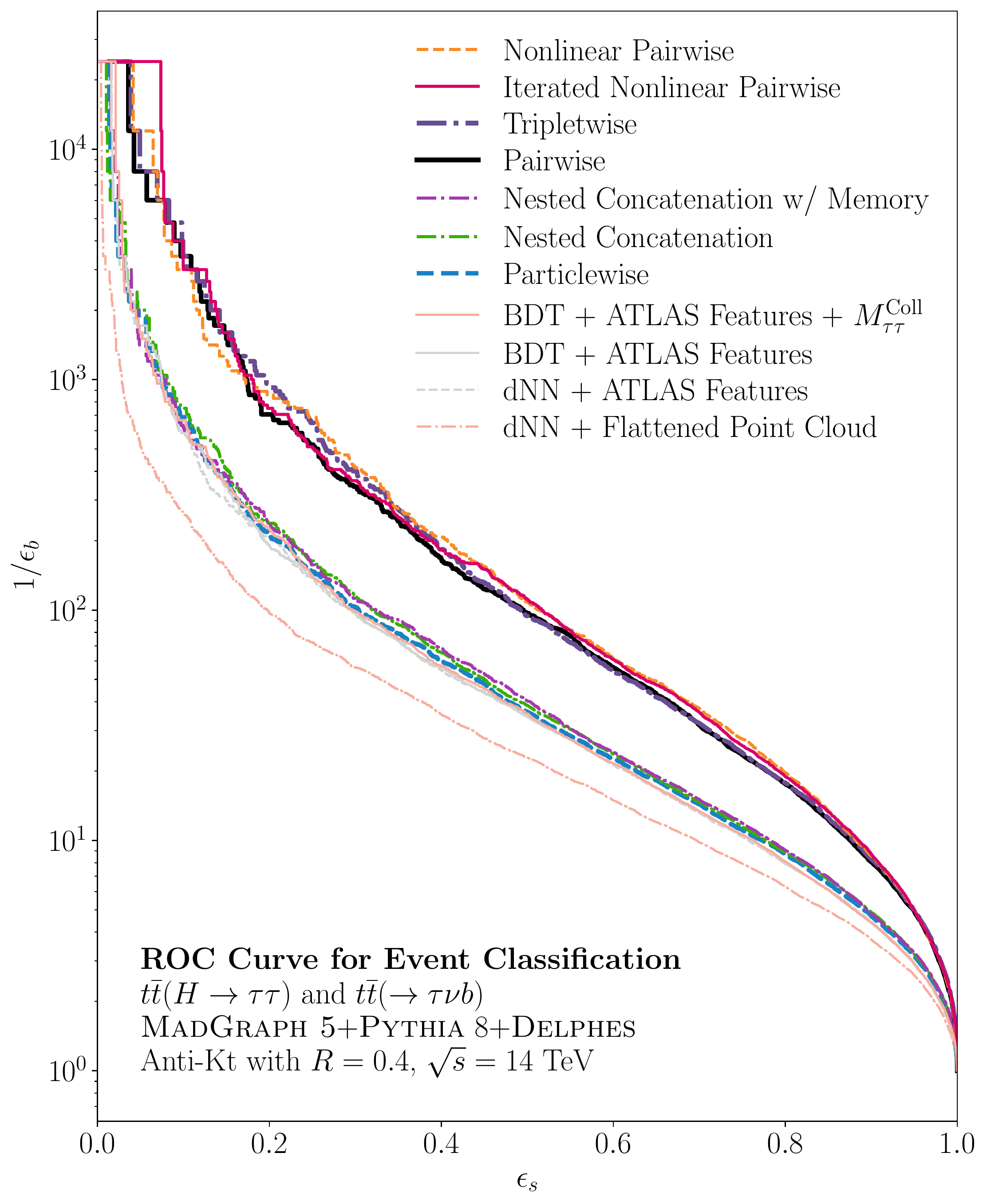} 
 \end{center} 
 \caption{
   ROC curves for event classification between $t\overline{t}H$ and $t\overline{t}$, comparing our proposed point cloud architectures to more traditional strategies.
 The signal efficiency $\epsilon_s$ is on the $x$-axis and the inverse background false-positive rate $1 / \epsilon_b$ is on the $y$-axis, such that better performance corresponds to curves that are more up and to the right.
 The Pairwise architecture, which is our recommended strategy, is one of the best performing methods for this task.
 }
 \label{fig:ROCs}
\end{figure}

\subsection{Performance of Point Cloud Architectures}
\label{sec:results}

\begin{table*}
  \caption{\label{tab:perf}
  Performance summary of the studied architectures.
  Here we tabulate: (1) the area under the ROC curve, (2) the number of trainable parameters, (3) the inverse false-positive rate $1 / \epsilon_b$ at fixed signal efficiency \(\epsilon_s=\{0.3,0.7\}\), which roughly corresponds to how many background events we see until a background event is mis-classified as a signal event, and (4) the ratio of signal efficiency to background false-positive rate $\epsilon_s / \epsilon_b$ at fixed signal efficiency \(\epsilon_s=\{0.3,0.7\}\), which measures the discriminatory power of the architecture. \neww{By comparing the average of the values in the $\epsilon_s=0.7$ columns of our best performing architectures (Nonlinear Pairwise, Iterated Nonlinear Pairwise, Tripletwise, and Pairiwse) to the best performing traditional architecture (BDT + ATLAS Features + $M_{\tau\tau}^{\rm Coll}$) we see that we gain about a 2.5 times increase in performance.} Bold-faced entries are the best performing in each column. 
  } 
     \begin{ruledtabular}
     \begin{tabular}{lc c c c c c }
    & & & \multicolumn{2}{c}{($\epsilon_s=0.3$)} & \multicolumn{2}{c}{($\epsilon_s=0.7$)}\\
    \cmidrule(lr){4-5}\cmidrule{6-7}
       Architecture &AUC&\# Params&$1/\epsilon_b$&$\epsilon_s / \epsilon_b$ &$1/\epsilon_b$&$\epsilon_s / \epsilon_b$ \\
    \hline
       Nonlinear Pairwise, \Eq{edgeconvnl}& \textbf{0.9590} & \textbf{96K} &\textbf{414.4} & \textbf{124.1} & \textbf{37.4} & \textbf{26.2}\\
       Iterated Nonlinear Pairwise, \Eq{iteratedecnl} & \textbf{0.9590} & 100K &364.2 & 109.2 & 35.9 & 25.2\\
 Tripletwise, \Eq{triplesum}& 0.9578 & 105K &400.6 & 120.1 & 32.3 & 22.6\\
 \textbf{Pairwise}, \Eq{pairwise_arch} & 0.9568 & 104K &343.4 & 102.3 & 31.7 & 22.2\\
Nested Concatenation w/ Memory, \Eq{gen_nested} & 0.9285 & 100K &111.8 & 33.5 & 15.1 & 10.6\\
 Nested Concatenation, \Eq{nested}& 0.9277 & 105K &116.1 & 34.7 & 14.6 & 10.2\\
 Particlewise, \Eq{particlewise_arch}& 0.9253 & 100K &102.7 & 30.8 & 14.4 & 10.1\\
\hline
BDT + ATLAS Features + $M_{\tau\tau}^{\rm Coll}$ & 0.9206 & - &98.5 & 29.5 & 13.4 & 9.4\\
BDT + ATLAS Features & 0.9201 & - &96.9 & 28.9 & 13.3 & 9.3\\
dNN + ATLAS Features & 0.9198 & 134K &99.3 & 29.6 & 13.2 & 9.2\\
dNN + Flattened Point Cloud & 0.9015 & 160K &56.3 & 16.9 & 9.8 & 6.9\\
    \end{tabular}
     \end{ruledtabular}

\end{table*}

We now compare the performance of our proposed point cloud architectures on the $t\overline{t}H$ versus $t\overline{t}$ event classification problem.
The point cloud architectures from \Sec{archs} that we test are:
\begin{itemize}
  \item Particlewise, \Eq{particlewise_arch};
  \item Pairwise, \Eq{pairwise_arch};
  \item Tripletwise, \Eq{triplesum};
  \item Nonlinear Pairwise, \Eq{edgeconvnl};
  \item Iterated Nonlinear Pairwise, \Eq{iteratedecnl};
  \item Nested Concatenation, \Eq{nested}; and
  \item Nested Concatenation with Memory, \Eq{gen_nested}.
\end{itemize}
The parameters for all architectures are summarized in \App{params}.
We chose hyperparameters such that each architecture has approximately 100k trainable parameters, to make sure we were comparing architectures based on their structure and not just on model size.
For comparison, we test four more traditional architectures:
 \begin{itemize}
  \item BDT trained with the ATLAS features;
  \item BDT trained with the ATLAS features and $M^{\rm Coll}_{\tau\tau}$;
  \item dNN trained with the ATLAS features; and
  \item dNN trained with the flattened point cloud,
\end{itemize}
where again the parameters are specified in \App{params}.

To assess the classification performance of each architecture, we plot their receiver operator characteristic (ROC) curves in \Fig{ROCs}.
These curves show the inverse background false-positive rate ($1/\epsilon_b$) as a function of the signal efficiency ($\epsilon_s$), as the cut on the architecture output is varied.
The best performing architectures are based on tripletwise or pairwise information.
The next best architectures use the nested concatenation structure from \Sec{global}.
The point cloud architecture with the weakest performance is the Particlewise architecture.
The dNN acting on the flattened point cloud has significantly worse performance, which indicates the importance of linking the point cloud data inputs to a suitable architecture for processing.

In \Tab{perf}, we tabulate the number of trainable parameters in each architecture, the area under the ROC curve (AUC), and various performance metrics at the operating points of $\epsilon_s=\{0.3,0,7\}$.
The choice of $\epsilon_s=0.7$ mimics the choice made in \RRef{ATLAS:2022yrq}.
At this operating point, our \neww{best performing models (Nonlinear Pairwise, Iterated Nonlinear Pairwise, Tripletwise, and Pairwise)} achieve a 2.5 times increase in $\epsilon_s / \epsilon_b$ over the more traditional dNN and BDT models.
This jump in performance would be even more pronounced if we choose an operating point of $\epsilon_s=0.3$, which leads to a nearly fourfold increase in discrimination power.%
\footnote{This operating point is also closer to the one that would maximize $\epsilon_s/\sqrt{\epsilon_b}$, i.e.\ the one that would yield the largest \emph{significance improvement} \cite{Gallicchio:2012ez}.}
Furthermore, the poor performance of the dNN on the flattened point cloud implies to us that the increase in performance in our architectures comes from their improved methods of processing information from a collider event and not from simply increasing the amount of information given to an architecture.

The processing of local information via tripletwise or pairwise features yields the most powerful classifiers, but this local information does not need to be processed in overly complex ways.
Notably, the architecture which processes information in nearly the simplest manner, the Pairwise architecture, is comparable to the performance of architectures with (iterated) non-linear structures.
This implies that more complex strategies to process local information do not necessarily lead to better performance, at least in this context of this event classification problem.

Finally, we note that machine learning architectures trained on Monte Carlo event samples are sensitive to simulation-specific behavior that may affect model performance on real data.
For example, the analysis in \RRef{ATLAS:2022yrq} did not utilize some input features because they were imperfectly modeled in simulation.
In our case study, though, some of these imperfectly modeled variables were used as inputs, so our architectures could be learning unphysical correlations.
Thus, the calibrated performance of our architectures on real data will likely be inferior compared to the performance on simulated data, unless some method is used to minimize the dependence on simulation-specific behavior.
Despite this caveat, we expect the \emph{relative} performance of the different point cloud strategies to be similar.

\subsection{Computational Cost of Point Cloud Architectures}

\begin{table}
  \caption{Computational cost summary of the studied architectures.
  Here, we tabulate the time per epoch and total number of epochs it took to train each model on a server equipped with two NVIDIA Tesla K80s.
  While training, we reserve 30\% of the training data as validation data and monitored validation loss.
  If validation loss has not improved in 32 consecutive epochs, we stop the training and restore the weights of the model to the point where validation loss was lowest.
  Bold-faced entries are the most computationally efficient in each column.
  }
\begin{ruledtabular}
  \begin{tabular}{lcc}
       &\multirow{2}*{\parbox{1.65cm}{Time/Epoch (seconds)}} & \multirow{2}*{\parbox{1.45cm}{Total \# of Epochs}}\\
       Architecture &&\\
    \hline
    
Tripletwise &240& 100\\
\textbf{Pairwise} &25& 93\\ 
       Nonlinear (NL) Pairwise &27& 88 \\
       Iterated NL Pairwise &50& 149\\
Nested Concatenation &20& 64\\
Nested Concatenation w/ Memory & 20& \textbf{54}\\
Particlewise &\textbf{10}& 61 \\
\hline
dNN + ATLAS Features &{10}& 110\\
dNN + Flattened Point Cloud&\textbf{5}& \textbf{38}\\
    \end{tabular}
     \end{ruledtabular}
  \label{tab:time}
\end{table}

Of the best performing architectures, the Pairwise architecture is the most computationally efficient.
To give an idea of the computational cost of our architectures, we tabulate the approximate time/epoch and total number of epochs to train each architecture in \Tab{time}.
These training times are obtained on a server equipped with two NVIDIA Tesla K80s.
The Pairwise architecture and Nonlinear Pairwise architecture are close in both performance and runtime efficiency.
The other two architectures that achieve similar performance, the Iterated Nonlinear and Tripletwise architectures, are significantly more computationally expensive.

Since the Pairwise architecture is nearly the best performing architecture while still being one of the simplest conceptually and most efficient computationally, we recommend the use of the Pairwise architecture for event classification.
For other point cloud tasks that have seen gains from using graph neural networks, we recommend their performance be benchmarked against the Pairwise architecture.

\subsection{{Latent Dimension Studies of the Pairwise Architecture}}
\label{sec:lat}

\begin{table*}
  \caption{\label{tab:lat}
  Impact of latent dimension size on the performance of the Pairwise architecture.
  We show the same performance metrics as \Tab{perf}, and the uncertainties correspond to the mean and variance from 8 random initializations.  
  Bold-faced entries are the best performing in each column.
  Even for a latent dimension size of $2^0 = 1$, the Pairwise architecture outperforms the baseline ATLAS method. 
}
     \begin{ruledtabular}
     \begin{tabular}{c c c c c cc}
    & & & \multicolumn{2}{c}{($\epsilon_s=0.3$)} & \multicolumn{2}{c}{($\epsilon_s=0.7$)}\\
    \cmidrule(lr){4-5}\cmidrule{6-7}
       Latent Dim.&AUC&\# Params&$1/\epsilon_b$&$\epsilon_s / \epsilon_b$ &$1/\epsilon_b$&$\epsilon_s / \epsilon_b$ \\
       \hline

$2^0$ & $0.9428 \pm 0.0025$ & 92K & $170.1\pm 10.2$ & $50.8\pm 3.0$ & $21.5\pm 1.4$ & $15.1\pm 1.0$\\
$2^1$ & $0.9452 \pm 0.0013$ & 92K & $205.1\pm 15.7$ & $61.3\pm 4.7$ & $22.7\pm 0.9$ & $15.9\pm 0.7$\\
$2^2$ & $0.9495 \pm 0.0022$ & 93K & $270.2\pm 38.6$ & $80.8\pm 11.5$ & $26.0\pm 1.8$ & $18.2\pm 1.2$\\
$2^3$ & $0.9526 \pm 0.0020$ & 94K & $326.8\pm 45.6$ & $97.6\pm 13.2$ & $29.2\pm 2.0$ & $20.4\pm 1.4$\\
$2^4$ & $0.9550 \pm 0.0010$ & 95K & $334.9\pm 23.2$ & $100.1\pm 6.9$ & $31.2\pm 1.6$ & $21.8\pm 1.1$\\
$2^5$ & $0.9565 \pm 0.0011$ & 98K & $358.0\pm 42.6$ & $106.5\pm 12.6$ & $32.8\pm 1.5$ & $23.0\pm 1.0$\\
$2^6$ & $0.9581 \pm 0.0006$ & 104K & $382.9\pm 37.8$ & $114.4\pm 11.4$ & $34.5\pm 1.4$ & $24.2\pm 1.0$\\
$2^7$ & $\mathbf{0.9584 \pm 0.0010}$ & 117K & $\mathbf{402.8\pm 43.5}$ & $\mathbf{120.7\pm 13.0}$ & $\mathbf{34.9\pm 1.4}$ & $\mathbf{24.4\pm 1.0}$\\

%
    \end{tabular}
     \end{ruledtabular}

\end{table*}
\begin{figure*}
  \subfloat[\label{fig:l2_nt}]{\includegraphics[width=0.5\linewidth]{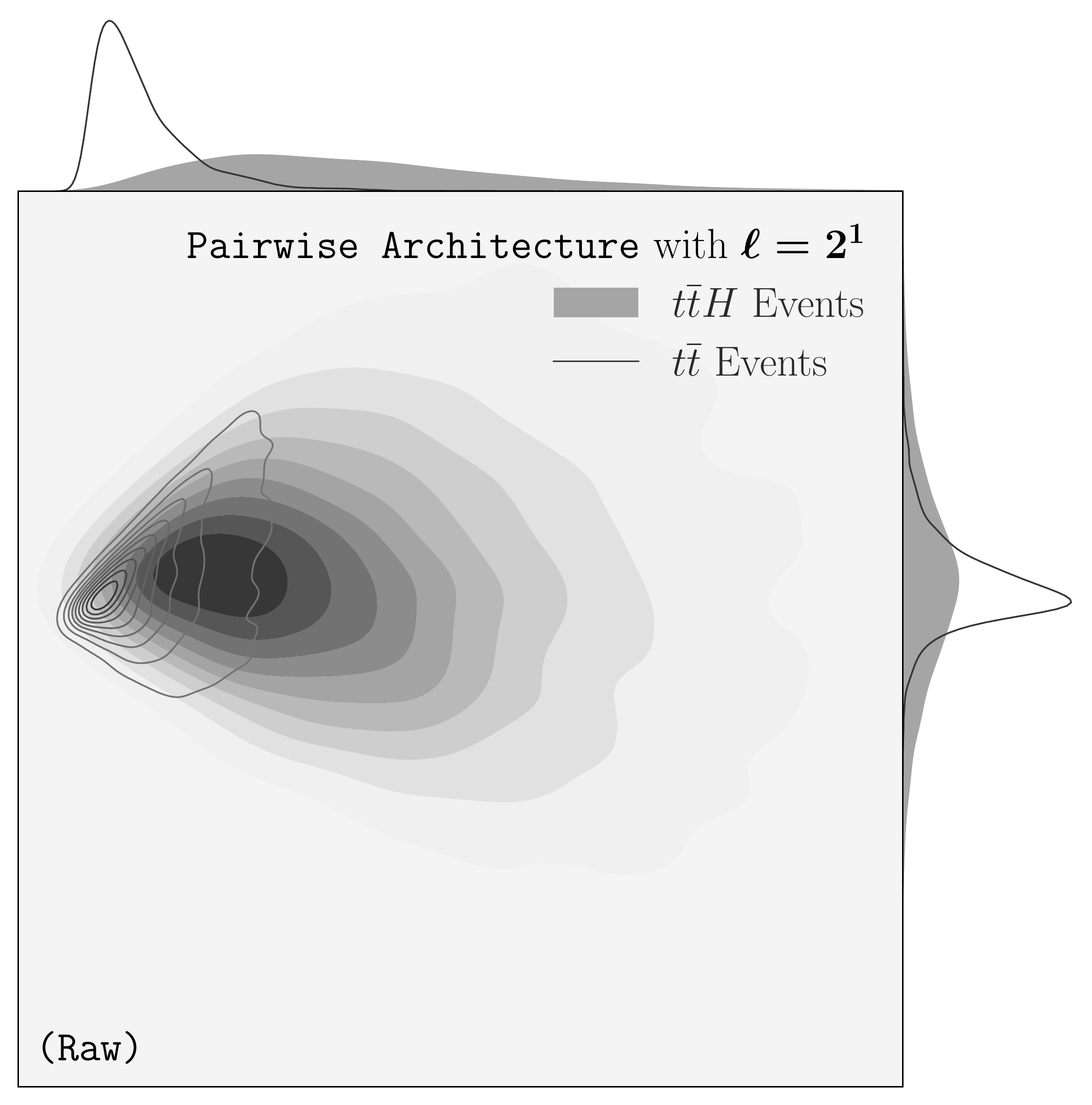}}
\subfloat[\label{fig:l2_t}]{\includegraphics[width=0.5\linewidth]{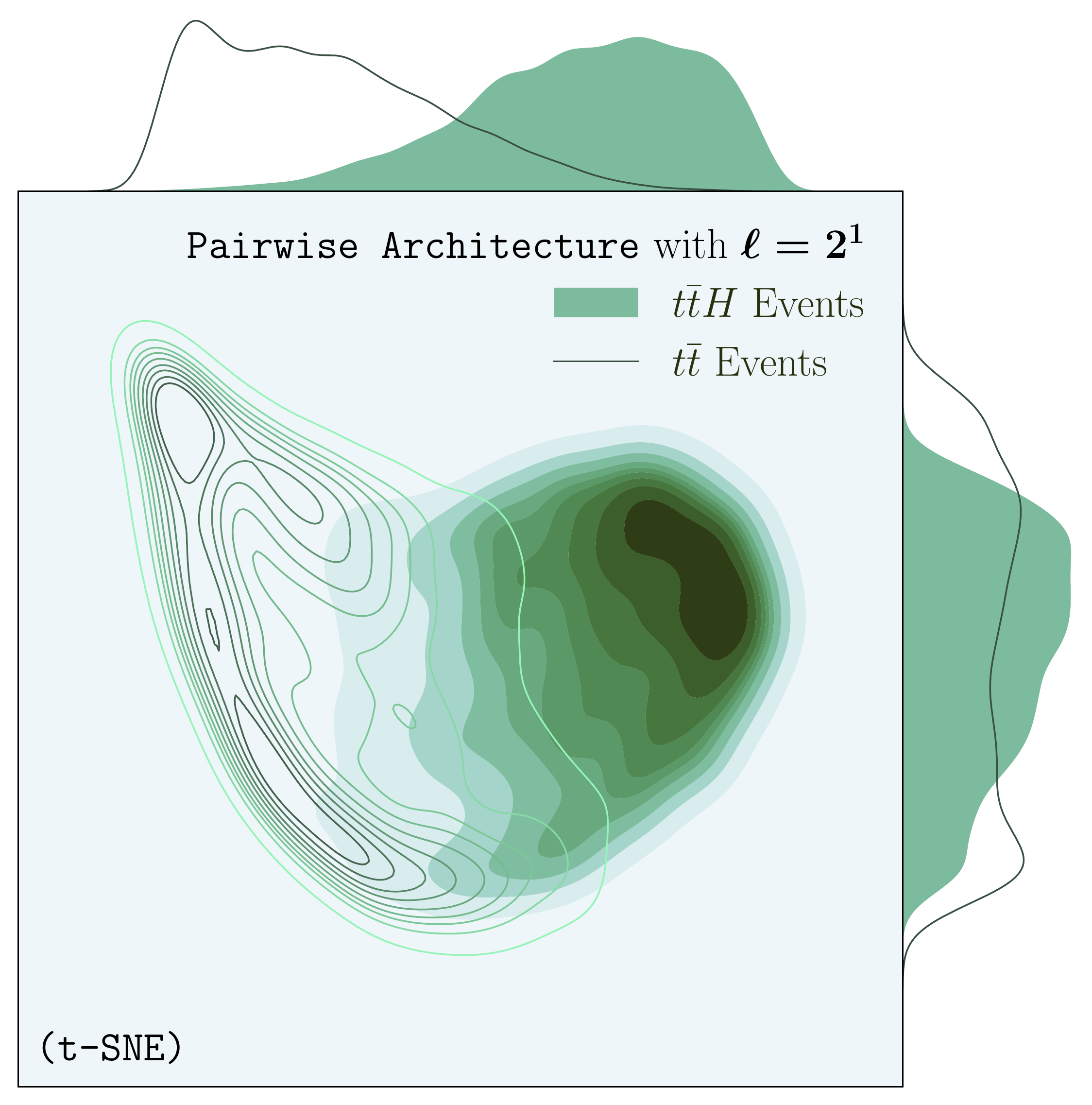}}
 \caption{ Visualization of the latent representations of events in the \(\ell=2^1\) Pairwise architecture, using KDE for density estimation.
   In (a), we plot the latent representations corresponding to the \(t\overline{t}H\) and \(t\overline{t}\) events.
 In (b), we apply t-SNE to disentangle the two distributions and see a clearer separation.
 Each plot also shows the marginalized distributions along each axis.}
 \label{fig:l2}
\end{figure*}

\begin{figure*}[h!]
  \subfloat[\label{fig:tatlas}]{\includegraphics[width=0.5\linewidth]{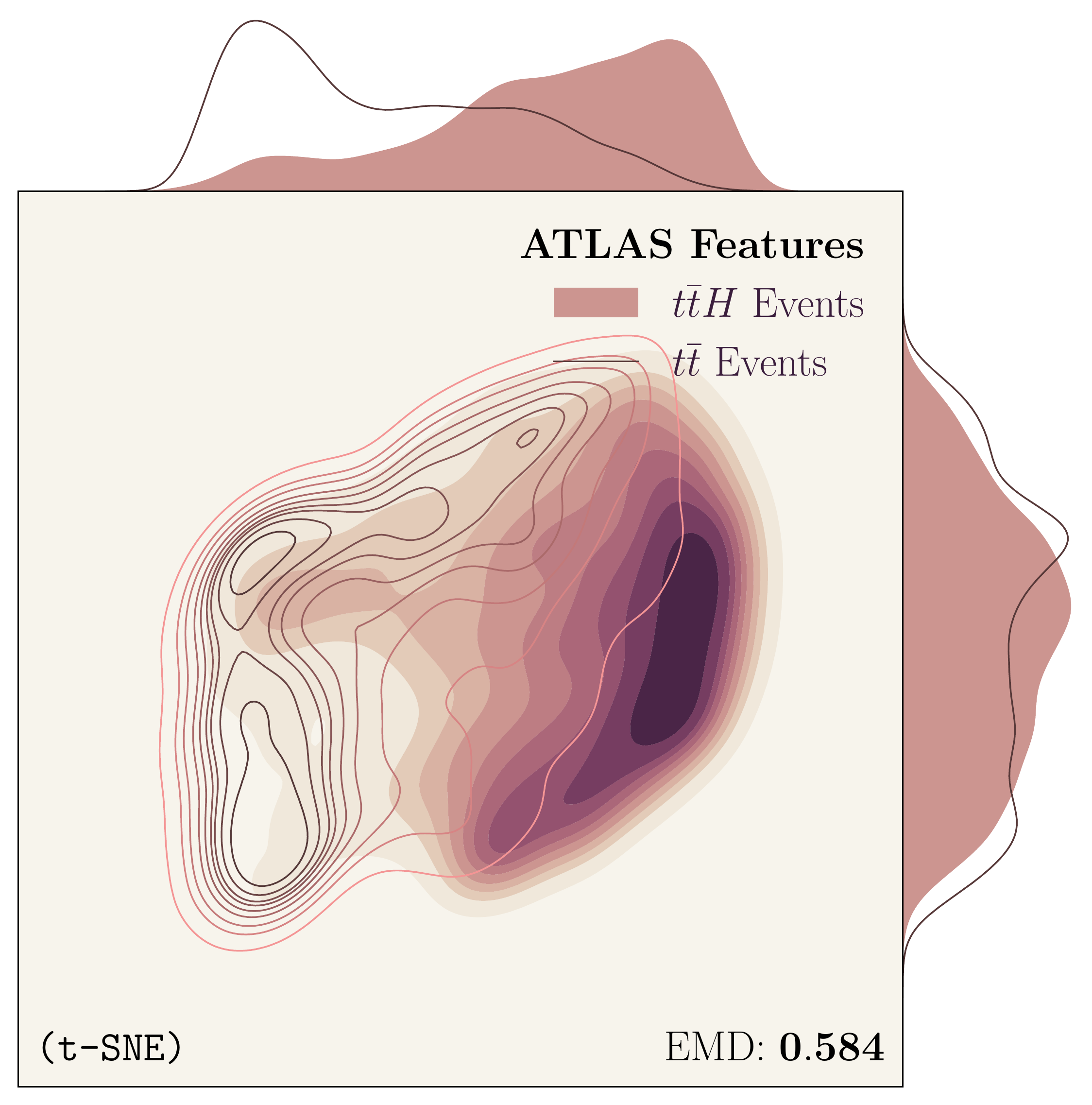}}
  \subfloat[\label{fig:tl2}]{\includegraphics[width=0.5\linewidth]{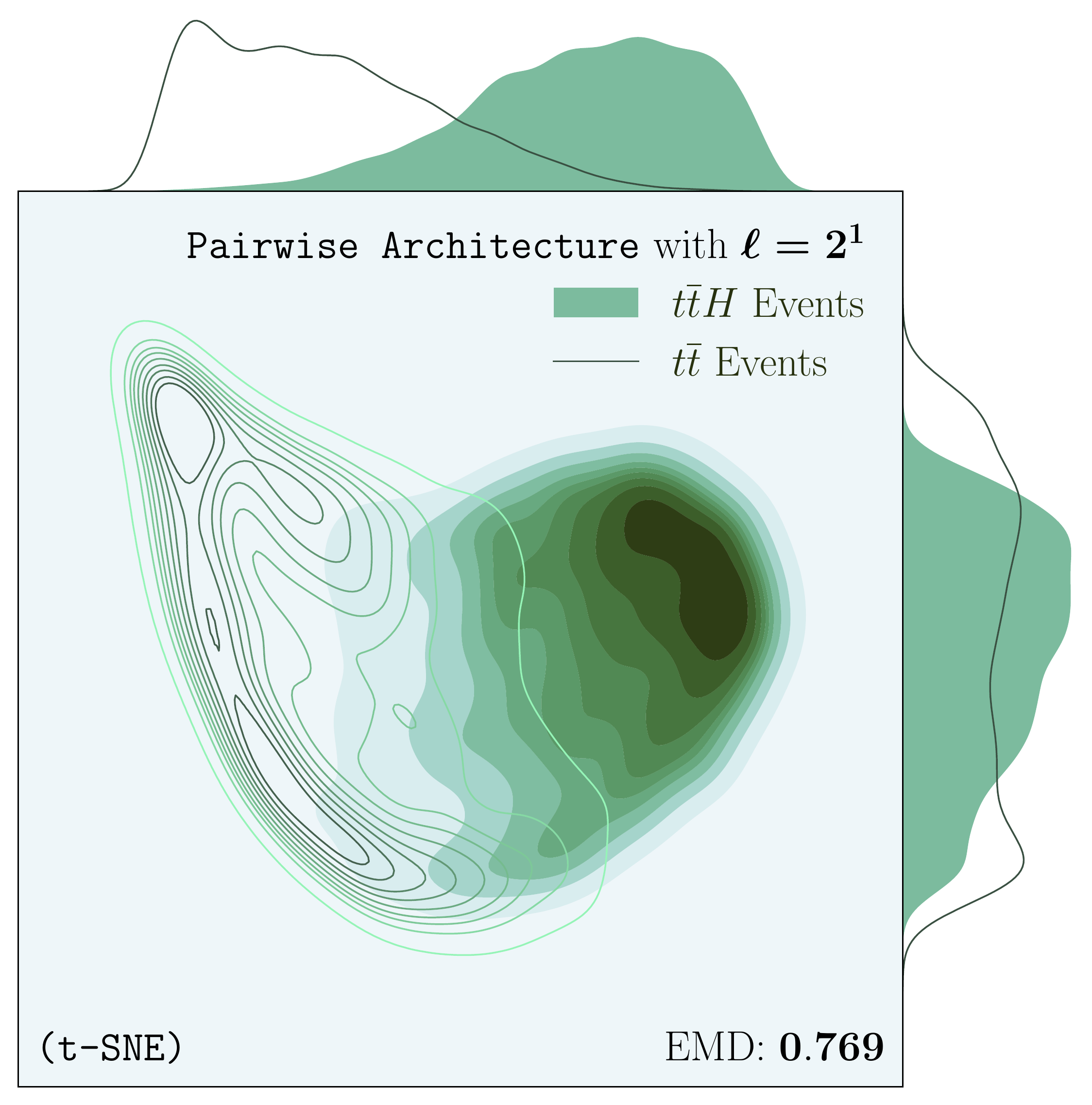}}\\
  \subfloat[\label{fig:tl8}]{\includegraphics[width=0.5\linewidth]{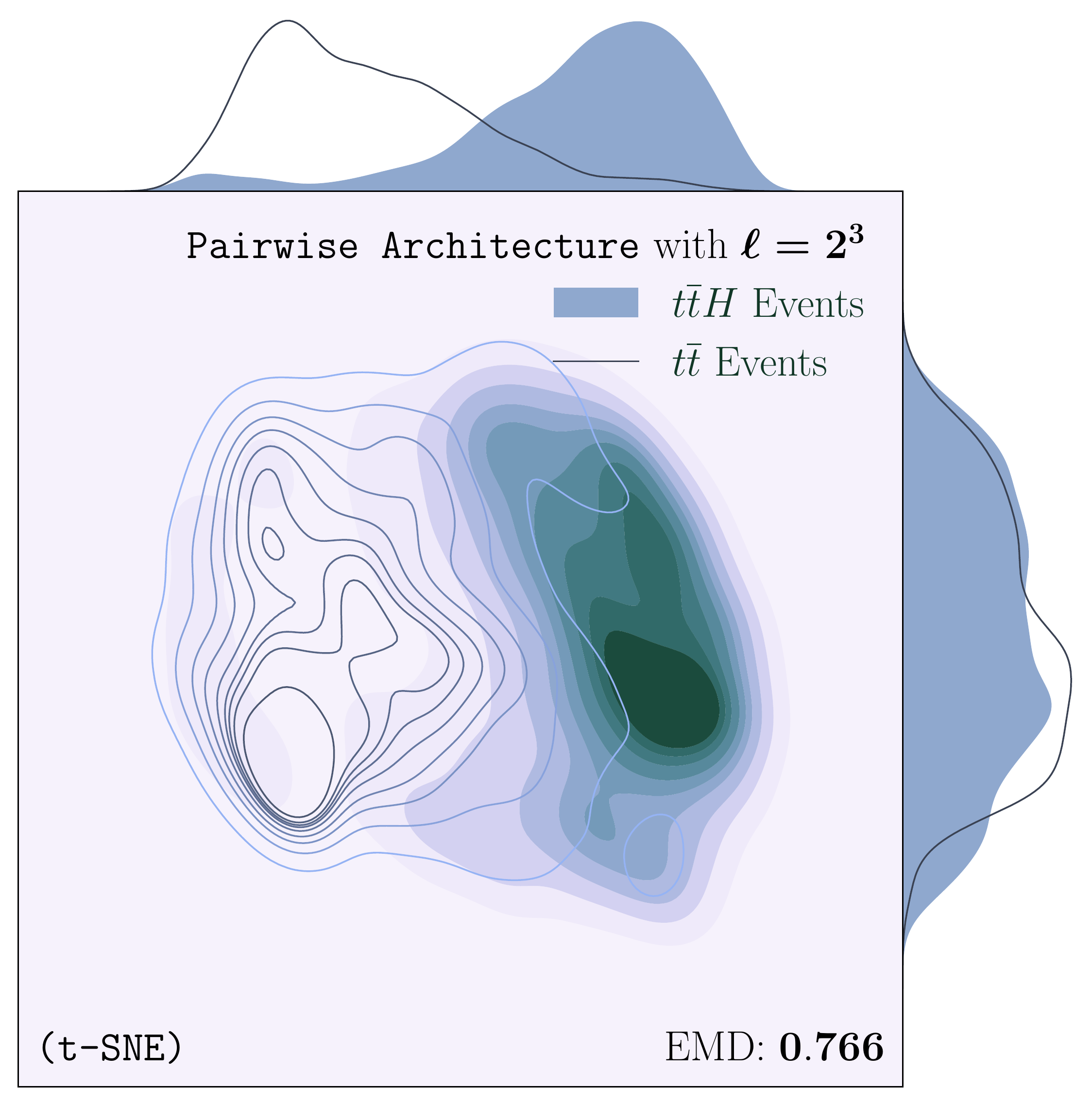}}
  \subfloat[\label{fig:tl64}]{\includegraphics[width=0.5\linewidth]{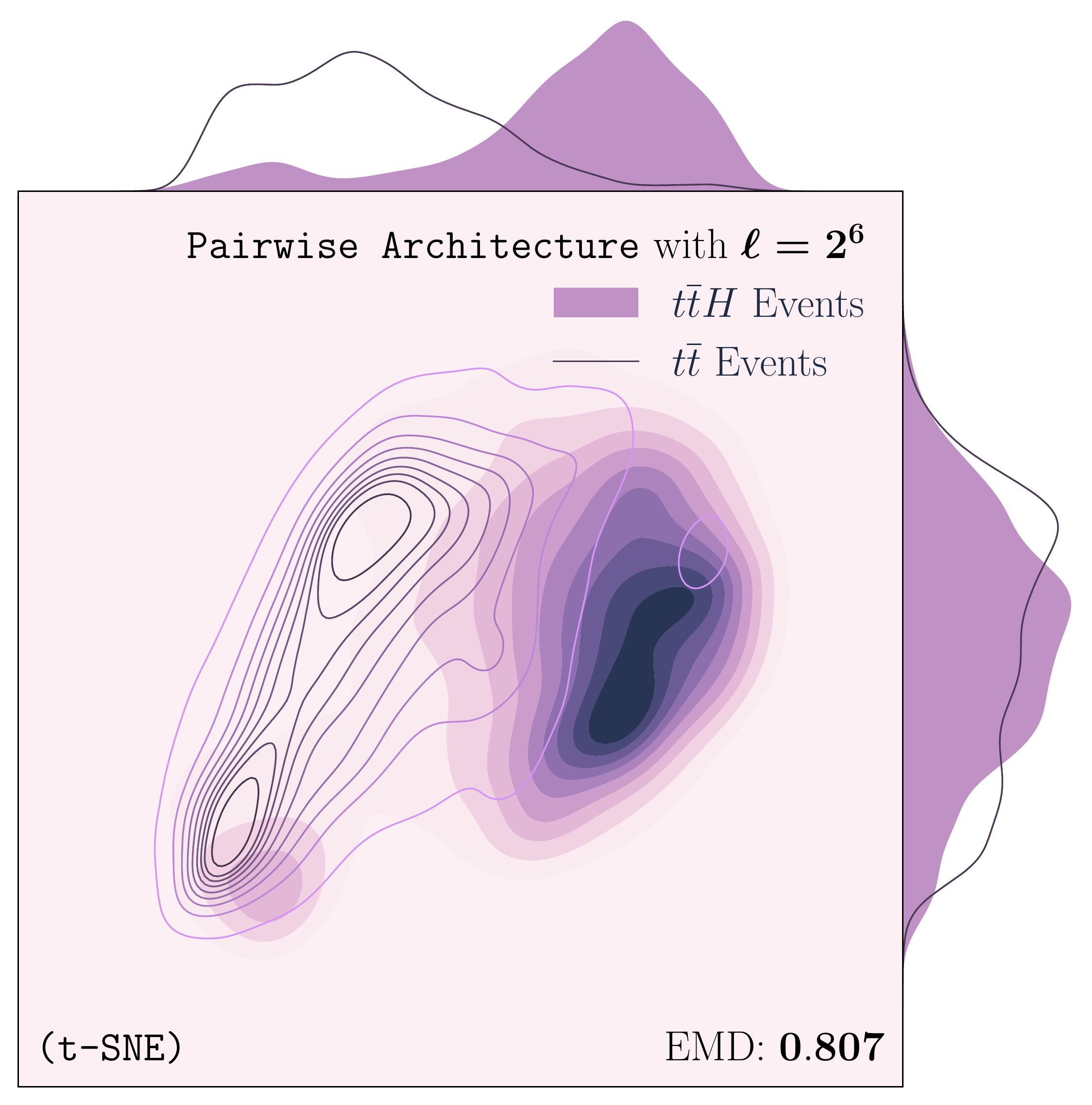}}
  \caption{ 
   Visualization of the latent representations using t-SNE embedding.
   Shown are
   (a) the ATLAS features which effectively have \(\ell=2^3\),
   (b) the Pairwise architecture with \(\ell=2^1\), which has the same information as \Fig{l2_t},
   (c) the Pairwise architecture with \(\ell=2^3\), and
   (d) the Pairwise architecture with \(\ell=2^6\).
   The t-SNE embeddings have been standardized such that the distributions have mean \(0\) and standard deviation \(1\) along both dimensions.
   The standardized embedding is then rotated such that the \(t\overline{t}H\) events are centered on the right side of the figure.
   For each plot, we report the EMD between the distribution of \(t\overline{t}H\) and \(t\overline{t}\) events, which roughly measures the separation of the two distributions, with larger EMD corresponding to better separation.
   We also plot marginalized densities along each axis.}
 \label{fig:latent}
\end{figure*}

Having convinced ourselves that the Pairwise architecture is a prime candidate for event classification, we now study the $\ell$-dimensional latent representation of events generated by this architecture.
Specifically, we study what happens if we restrict the latent dimension $\ell$ of the Pairwise architecture.
This gives us a way to study how powerful this architecture is at identifying useful discriminatory features from the event kinematics.
Furthermore, we can visualize the latent representations to get some picture of what the architecture is physically learning.

The latent dimension $\ell$ corresponds to the number of discriminatory features the architecture can extract from the point cloud.
Concretely, if we restrict the latent dimension of the Pairwise architecture to $\ell=2$, we are essentially asking the architecture to extract two features from the point cloud that, when processed by the $F$ function, can robustly distinguish between signal and background events.
In the Pairwise architecture from \Eq{pairwise_arch}, the latent representation of an event $\mathcal X$ is the result of a double summation over pairs of particles:  $\sum_{i,j}\Phi_2(x_i,x_j)\in\mathbb R^\ell$.
In \Fig{ROCs} and  \Tab{perf} above, we used $\ell=2^6$, as described in \App{params}.
We can think of the BDTs and dNNs trained on the ATLAS variables~\cite{ATLAS:2022yrq} as having $\ell=2^3$, because they take as input $8$ hand-engineered features, as described in \Sec{process}.

As shown in \Tab{lat}, we can significantly restrict the size of the latent dimension of our Pairwise architecture and still outperform traditional methods of event classification.
This table shows the performance of the Pairwise architecture as a function of the size of the latent dimension.
The strong performance even for $\ell=2^0 = 1$ implies that deep-learning-driven feature engineering is extremely powerful for finding robust discriminatory features for event classification.
These discriminatory features could in principle be found from the traditional feature engineering game, but we expect they would be extremely difficult to find in practice without the use of machine learning.

\begin{figure*}
 \begin{center}
 \includegraphics[width=\linewidth]{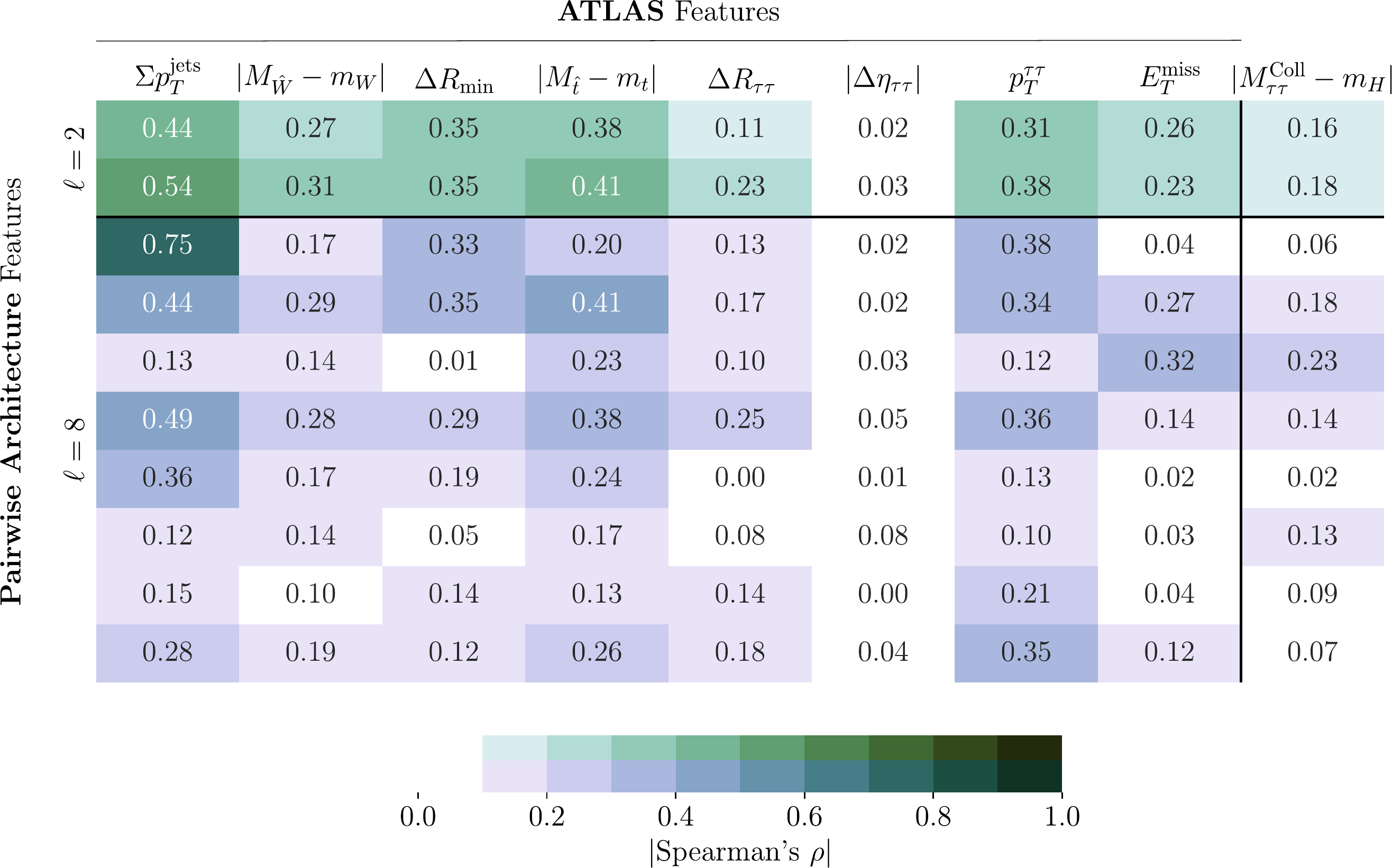}
 \end{center}
 \caption{
   The Spearman's rank correlation coefficients of the ATLAS features from \RRef{ATLAS:2022yrq} (left block) and ditau mass (right block) with the latent features learned by our Pairwise architectures, using $\ell=2$ (top block) and $\ell=8$ (bottom block).
   Here, \(m_W=80.4\) GeV, \(m_t=172.5\) GeV, and \(m_H=125\) GeV.
   We compute the \(\rho\) between the Pairwise architecture feature and the absolute difference of \(M_{\hat{W}}\), \(M_{\hat{t}}\), and \(M_{\tau\tau}\) to canonical values, since the ``goodness'' of these mass variables is monotonic with how close these variables are to the true \(W\)-boson, top quark, and Higgs mass.
   Of all the chosen ATLAS features, the scalar sum of the jets' transverse momenta is particularly correlated with the features used by the Pairwise architecture.
   Also noteworthy is the presence of a correlation between \(M^{\rm Coll}_{\tau\tau}\) and the Pairwise latent features.
 }
 \label{fig:spearman} 
\end{figure*}
\begin{figure}[t]
   \includegraphics[width=\linewidth]{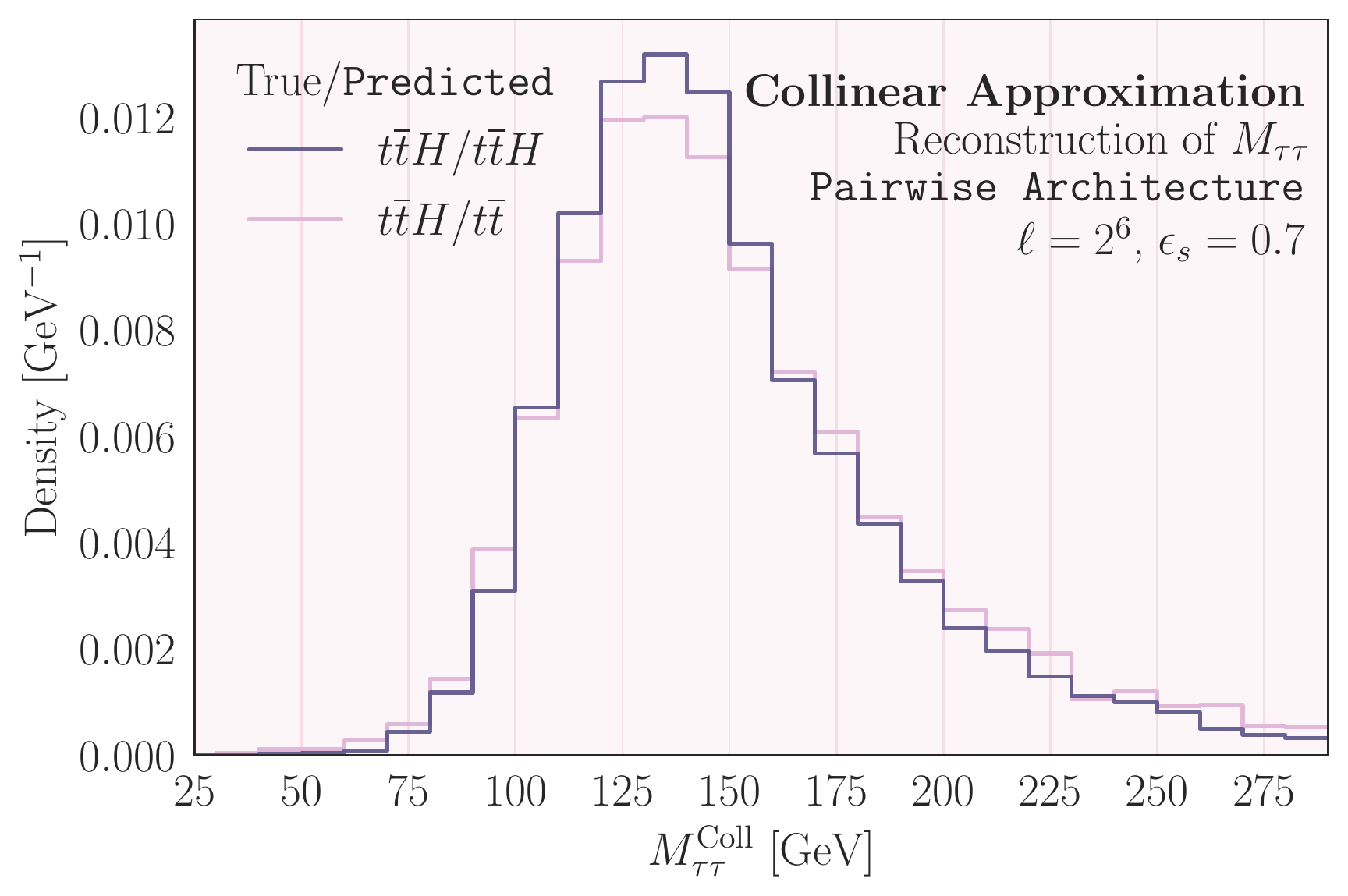}
 \caption{
Normalized distributions of $M^{\rm Coll}_{\tau\tau}$ for  $t\overline{t}H$ signal events 
Predicted labels are from the Pairwise architecture with $\ell=2^6$ at a fixed signal efficiency of $\epsilon_s=0.7$, where the purple curves correspond to $t\overline{t}H$-labeled events and the pink curves correspond to $t\overline{t}$-labeled events.
Events classified as $t\overline{t}H$ have a sharper $M^{\rm Coll}_{\tau\tau}$ feature, suggesting that the Pairwise architecture has learned features with some correlation to $M^{\rm Coll}_{\tau\tau}$.
 }

\label{fig:mtt1}
\end{figure}

For $\ell = 2$, we can directly visualize the latent space of the Pairwise architecture, as shown in \Fig{l2}.
Here we plot two versions of the latent dimension, where densities are approximated with kernel density estimation (KDE) \cite{Rosenblatt1956, Parzen1962}.%
\footnote{Due to computational restrictions, we generate the figures in this subsection using roughly a third of the events from the testing set.}
In \Fig{l2_nt}, we plot the densities of the raw latent representations.
We see that one of the latent space variables yields a clean separation between signal and background events, while the other one yields an approximately Gaussian distributed feature with different widths for signal and background.
In \Fig{l2_t}, we apply t-distributed stochastic neighbor embedding (t-SNE) \cite{VanDerMaaten2009,VanDerMaaten2012,Garcia-Alonso2014,VanDerMaaten2015} to the latent representation, which shows more clearly the separation between signal and background events.
This t-SNE visualization serves as a reference for later plots with higher $\ell$.

As we increase the latent dimension $\ell$, the trained architectures identify more discriminatory features, leading to better separation between signal and background events in the latent space.
In \Fig{latent}, we plot the 2-dimensional t-SNE embeddings of the 8-dimensional ATLAS feature space and compare it to our $\ell=\{2^1, 2^3,2^6\}$ Pairwise architectures.
To quantify the separation between the two (t-SNE projected) distributions, we approximate the Earth Mover's Distance (EMD) \cite{Peleg1989, Rubner, Rubner2000} using the Euclidean distance as the ground metric between the $t\overline{t}H$ and $t\overline{t}$ distributions.
Since the length scales within a t-SNE embedding are not physical and we wish for the EMD to be a meaningful metric of comparison, we standardize the whole distribution of events in each plot so that along each dimension we have zero mean and unit variance.
Qualitatively, we see that for our proposed Pairwise architecture, the joint and marginalized distributions of $t\overline{t}H$ and $t\overline{t}$ are more clearly separated than for the ATLAS features.
This observation is reinforced quantitatively by seeing that the EMD between the embedded distribution of signal and background events is smallest for the ATLAS features, implying the largest degree of overlap.

\subsection{Correlations between Pairwise Architecture Features and Hand-Engineered Features}

The features found to be useful to represent collider events by our Pairwise architecture are correlated to several of the features chosen by ATLAS to classify events.
In \Fig{spearman}, we tabulate the Spearman's rank correlation coefficients (Spearman's $\rho$) between the ATLAS features used in \RRef{ATLAS:2022yrq} (see \Sec{process}) and the learned latent features of our Pairwise architecture with $\ell=2$ and $\ell=8$. 
Spearman's $\rho$ quantifies the degree to which two variables are monotonically related, with $\rho = 0$ meaning no correlation and $\rho =1$ meaning perfect correlation.
Unlike the more common Pearson correlation coefficient, Spearman's $\rho$ does not care if the relationship is linear or not, which makes it a more robust notion of correlation in the context of non-linear neural networks.

Because Spearman's $\rho$ quantifies monotonic relations, we have to be mindful of features whose classification performance is related to how close they are to the true $W$ boson mass ($m_W = 80.4$~GeV), top quark mass ($m_t = 172.5$~GeV), and Higgs boson mass ($m_H = 125$~GeV).
Thus, instead of tabulating correlations for $M_{\hat{W}}$, $M_{\hat{t}}$, and $M^{\rm Coll}_{\tau \tau}$, we look at $|M_{\hat{W}} - m_W|$, $|M_{\hat{t}}-m_t|$, and $|M^{\rm Coll}_{\tau\tau}-m_H|$.
We see that of all the ATLAS features, the scalar sum of all jet \(p^T\) is most closely related to the Pairwise architecture features.
Other ATLAS features that have some correlation with the Pairwise architecture features are the smallest $\Delta R$ between two jets, the $p^T$ of the \(\tau\tau\) dijet, the $\Delta R$ between the two \(\tau\)-tagged jets, and the invariant mass of the dijet/trijet with invariant mass closest to the $W$ boson/top quark mass.

From \Fig{spearman}, we can see that there exists a correlation between $M^{\rm Coll}_{\tau\tau}$ and the Pairwise architecture features, as anticipated in \Sec{process}. 
This correlation is subtle, though, and not captured by a single latent space feature.
In \Fig{mtt1}, we consider the Pairwise architecture with \(\ell=2^6\), and plot the $M^{\rm Coll}_{\tau\tau}$ distribution for \(t\overline{t}H\) events correctly and incorrectly classified by this architecture at fixed signal efficiency \(\epsilon_s=0.7\).
We see that correctly labeled events have a sharper peak in $M^{\rm Coll}_{\tau\tau}$, albeit shifted a bit to the right of $m_H$.
This behavior suggests that the Pairwise architecture with \(\ell=2^6\), which was used for our comparison comparison in \Sec{results}, has learned features with some correlation to $M^{\rm Coll}_{\tau\tau}$.

\section{Conclusions}
\label{sec:conclusion}
\ptitle{Conclusion}

In this paper, we compared neural network architectures defined on the point cloud representation of collider events with traditional approaches for event classification.
The point cloud representation allows us to circumvent many difficulties arising from trying to robustly represent an event with a fixed-size input. 
Our architectures explore three complementary strategies to process information within a point cloud:
(1) using multiple summations to improve local information processing;
(2) using iterated convolutions to increase an architecture's power to build latent representations; and
(3) using nested concatenation of global features to improve global information processing.
These can be viewed as simplified versions of the strategies used to build graph neural networks.

To benchmark our architectures, we performed a case study of event classification in the $H\rightarrow\tau\tau$ channel and compared the results to an ATLAS study using hand-engineered features~\cite{ATLAS:2022yrq}.
At a comparable signal efficiency operating point to the one used by ATLAS, we found a 2.5 times increase in background rejection.
This gain in performance was not simply due to the increased size of the input space.
Indeed, when the flattened point cloud representation was processed with a dNN, we found worse performance than for the ATLAS baseline.
We therefore recommend further explorations of point cloud architectures for event classification problems.

Among the tested architectures, the Pairwise architecture exhibited the best balance of classification performance, computationally efficiency, and conceptual simplicity.
The Particlewise architecture, based on a straightforward application of the Deep Sets formalism~\cite{Zaheer2017}, yielded performance similar to the ATLAS baseline.
By considering the set of \emph{pairs} of particles, we found a boost in performance without requiring more complex nonlinear or iterated structures as in EdgeConv architectures~\cite{Wang:2018nkf}.
The Pairwise architecture continues to have good discriminatory power as the latent space dimension $\ell$ is decreased, and by visualizing the learned latent representations, we conclude that the Pairwise architecture is able to identify discriminatory features that are well suited for event classification.
We found that these learned features are correlated with traditionally chosen hand-crafted features.
While more complex graph neural networks might provide better performance for certain event-wide tasks, we recommend that they be benchmarked against the simpler Pairwise architecture.

\neww{
A key open question regarding our proposed architectures is how they will perform as the final state multiplicity increases. 
In this paper, we considered a final state of $O(10)$ objects, but final states of interest with $O(100)$ or even $O(1000)$ objects also appear in particle physics applications such as jet substructure studies. 
We found that for our case of $O(10)$ final state objects, the pairwise architecture achieved the best balance between performance and cost. 
As we scale up to more final state objects, however, it will be important to understand the performance/cost balance, as pairwise architectures scale quadratically with the number of objects being studied. 
To extend our architectures to more final state objects without suffering from quadratic scaling, one could follow the strategy of the original EdgeConv construction \cite{Wang:2018nkf} and consider only $k$-nearest neighbors instead of all pairs. 
These trade-offs are an important area for future studies.
} 

There are several more directions for further explorations.
First, our case study focused on binary classification, but as described in \Eq{general_notation}, these point cloud architectures could be applied to multi-category classification or regression.
Second, the Pairwise architecture is based on learning a generic function $\Phi_2$, but it may be possible to improve performance and interpretability by restricting its functional form.
Finally, there has been a rising interest in incorporating physical symmetries into neural networks.
While our point cloud architectures already exhibit manifest permutation invariance among the particles, event classification could benefit from directly incorporating Lorentz symmetry \cite{Butter:2017cot, Erdmann:2018shi, Bogatskiy:2020tje, Gong:2022lye, Bogatskiy:2022hub, Qiu:2022xvr} or infrared and collinear safety \cite{Komiske:2018cqr, Dolan:2020qkr, Chakraborty:2019imr, Chakraborty:2020yfc, Konar:2021zdg, Romero:2021qlf, Atkinson:2022uzb}.

\begin{acknowledgments}
J.T. thanks Patrick Komiske, Eric Metodiev, Nishat Protyasha, and Raymond Wynne for collaboration on unpublished work related to alternative point cloud architectures. 
D.S. and P.O. were supported by the US Department of Energy, Office of Science, Office of High Energy Physics, under Award Number DE-SC0007890.
J.T. is supported by the U.S. Department of Energy (DOE) Office of High Energy Physics under Grant No. DE-SC0012567, and by the National Science Foundation under Cooperative Agreement PHY-2019786 (The NSF AI Institute for Artificial Intelligence and Fundamental Interactions, \url{http://iaifi.org/}). 
\end{acknowledgments}

\appendix

\section{Permutation Equivariant Deep Sets}
\label{app:equiv_deep_set}

In \Sec{global}, we argued that concatenation (with or without memory) is a powerful way to incorporate global information into local particle processing.
An alternative additive approach to incorporating global information was presented in \reft{Zaheer2017} and applied in \RRef{Dolan:2020qkr}.
In this appendix, we show that the additive approach can be viewed as a special case of the concatenation approach with memory.

A permutation equivariant Deep Set that maps ${\color{c1}n}$-dimensional point clouds to ${\color{c3}\ell}$-dimensional point clouds can be written as follows~\cite{Zaheer2017}:
      \begin{equation}
        \Omega(x_i, \mathcal X) = \sigma(x_i \, \vect\Lambda- \hat g(\mathcal X ) \, \vect\Gamma).
      \label{eq:equiv_deep_set}
    \end{equation}
Here, $\sigma$ is some activation function, $\hat{g}(\mathcal X)\in\mathbb R^{1\times {\color{c1}n}}$ is the result of a symmetric aggregation operation over the point cloud, $\vect\Lambda$ is a transformation of the features of particle $x_i\in\mathbb R^{1\times {\color{c1}n}}$, and $\vect\Gamma\in\mathbb R^{{\color{c1}n}\times \ell}$ is a transformation of the pooled features.
One can iteratively apply these equivariant Deep Set layers to create a permutation equivariant architecture, which was shown to be universal for permutation equivariant functions in \reft{Segol2020}.

We now show that a single permutation equivariant Deep Sets layer from \Eq{equiv_deep_set}  is a special case of the concatenation approach from \Eq{concat_for_AppA}.
First, since we are looking at one layer, we can take
\begin{equation}
\mathcal X^{(d-1)} \equiv \mathcal X.
\end{equation}
Since Deep Sets are universal for permutation invariant functions and $\hat{g}$ is by definition permutation invariant, we can assume that
\begin{equation}
f^{(d-1)}(\mathcal X^{(d-1)}) \equiv \hat{g}(\mathcal X).
\end{equation}
Now consider implementing $\Phi^{(d)}$ with a one-layer neural network with the same activation function $\sigma$ as in \Eq{equiv_deep_set} and weights \(\vect w^{(d)}\in\mathbb R^{2{\color{c1}n}\times {\color{c3}\ell}}\).
With these assumptions, \Eq{concat_for_AppA} reduces to 
\begin{align}
  \nonumber
  x_i^{(d)} &= \Phi^{(d)}\left( x_i^{(d-1)}\oplus f^{(d-1)} \left( \mathcal X^{(d-1)} \right)  \right) \\
            &= \sigma \left( \left( x_i^{(d-1)}\oplus \hat{g}(\mathcal X)  \right) \vect w^{(d)} \right) 
            \label{eq:interm_in_AppA}.
\end{align}
Since we are multiplying the concatenation of two vectors with the weight matrix $\vect w^{(d)}$, we can decompose the weight matrix as
\begin{equation}
  \vect w^{(d)} = \begin{bmatrix} \vect \lambda^{(d)}\\\vect\gamma^{(d)} \end{bmatrix}, \qquad \vect\lambda^{(d)},\vect\gamma^{(d)} \in \mathbb R^{{\color{c1}n},{\color{c3}\ell}}.
\end{equation}
This reduces \Eq{interm_in_AppA} to 
\begin{equation}
  x_i^{(d)} = \sigma \left( x_i^{(d-1)}\vect\lambda^{(d)} + \hat{g}(\mathcal X) \, \vect \gamma^{(d)} \right).
  \label{eq:interm2_in_AppA}
\end{equation}
Finally, if we choose \(\vect\lambda^{(d)} = \vect\Lambda\)  and \(\vect \gamma^{(d)} = -\vect\Gamma\) then \Eq{interm2_in_AppA} is equivalent to \Eq{equiv_deep_set}.

We therefore conclude that a single permutation equivariant Deep Set layer is a special case of the concatenation approach.
This in turn means that the composition of $L$-equivariant Deep Set layers from \Eq{equiv_deep_set} is equivalent to the resulting point cloud after $L$-nested layers of \Eq{concat_for_AppA}.
Thus, the architecture in \reft{Zaheer2017} based on iteratively applied equivariant Deep Set layers is a special case of our Nested Concatenation with Memory architecture.

\section{Model Parameters}
\label{app:params}

In this appendix, we specify the model parameters used for the event classification study in \Sec{results}.
All architectures were implemented with {\sc Keras} \cite{chollet2015keras} using the {\sc TensorFlow} \cite{Abadi:2016kic} back-end.
We chose model parameters so that the total number of trainable parameters in each model was roughly the same across all architectures.

\begin{itemize}
  \item \textbf{Particlewise Architecture, \Eq{inv_deep_set}:} 
    \begin{itemize}
      \item \(F\):  4 layers each 128 nodes wide
      \item \(\Phi\):  4 layers with widths (128, 128, 128, 64)
    \end{itemize}
  \item \textbf{Pairwise Architecture, \Eq{pairwise_arch}}: 
    \begin{itemize}
      \item \(F\): 5 layers each 64 nodes wide
      \item \(\Phi\): 5 layers with widths (64, 128, 256, 128, 64)
    \end{itemize}
  \item \textbf{Tripletwise Architecture, \Eq{triplesum}:}\begin{itemize}
    \item \(F\):  5 layers each \(64\) nodes wide
    \item \(\Phi\):  5 layers with widths (64, 128, 256, 128, 64)
  \end{itemize}
    \item \textbf{Nonlinear Pairwise Architecture, \Eq{edgeconvnl}:}
      \begin{itemize}
        \item \(F\):  5 layers each 32 nodes wide
        \item \(\Phi\):  5 layers with widths (64, 128, 256, 128, 64)
        \item \(\Pi\):  3 layers each 32 nodes wide
      \end{itemize}
    \item \textbf{Iterated Nonlinear Pairwise Architecture, \Eq{iteratedecnl}}: 
      \begin{itemize}
        \item \(L\): Since we found that performance does not improve for \(L>3\), we use \(L=3\)
        \item \(F\): 5 layers each 32 nodes wide
        \item \(\Pi^{(i)}\): All have 3 layers each 32 nodes wide
        \item \(\Phi^{(i)}\): All have \(5\) layers with widths (64, 64, 116, 64, 64)
      \end{itemize}
    \item \textbf{Nested Concatenation Architecture, \Eq{nested}:} 
      \begin{itemize}
        \item \(L\): Since we found that performance does not improve for \(L>2\), we use \(L=2\)
        \item \(F^{(i)}\): All have 4 layers each 70 nodes wide
        \item \(\Phi^{(i)}\): For \(i\ne L\), the \(\Phi^{(i)}\) have 3 layers each 70 nodes wide; for \(i=L\), \(\Phi^{(L)}\) has 4 layers with widths (70, 70, 70, 64)
      \end{itemize}    
    \item \textbf{Nested Concatenation w/ Memory Architecture, \Eq{gen_nested}:} 
      \begin{itemize}
        \item \(L\): Since we found that performance does not improve for \(L>2\) we use \(L=2\)
        \item \(F^{(i)}\): All have 3 layers each 68 nodes wide
        \item \(\Phi^{(i)}\): For \(i\ne L\), the \(\Phi^{(i)}\) have 3 layers each 68 nodes wide; for \(i=L\), \(\Phi^{(L)}\) has 4 layers with widths (68, 68, 68, 64)
      \end{itemize}
    \item \textbf{Dense Neural Network:} (ATLAS/Naive Features)
      \begin{itemize}
        \item We implement dNNs with a batch normalization layer \cite{Ioffe:2015ovl} followed by 3 layers each 256 nodes wide. 
      \end{itemize}
\end{itemize}

We use \texttt{leaky\_relu} activation functions between each layer in all of our neural networks.
A two-unit layer followed by a \texttt{softmax} activation function is used as the output layer in all models.
To train, we minimize categorical cross entropy using the {\sc Adam} optimization algorithm \cite{Kingma:2014vow} with the {\sc AMSGrad} enhancement \cite{Reddi2019}.
When training, we reserve \(30\%\) of the training data as validation data and monitor the validation loss.
To avoid over-fitting, we stop training if validation loss has not improved in 32 epochs and restore the weights of the model to the point when the validation loss was lowest.

The following BDT parameters were used in our model implemented with {\sc XGBoost} \cite{Chen:2016:XST:2939672.2939785}.
These parameter were chosen using the hyperparameter tuning library {\sc Hyperopt} \cite{Bergstra2013} with the Tree of Parzen Estimators algorithm \cite{Bergstra2011}.
\begin{itemize}
 \item \texttt{colsample\_bytree}: 0.703
 \item \texttt{eta}: 0.35
 \item \texttt{gamma}: 1.02
 \item \texttt{max\_depth}: 10
 \item \texttt{min\_child\_weight}: 10
 \item \texttt{n\_estimators}: 297
 \item \texttt{reg\_alpha}: 40.0
 \item \texttt{reg\_lambda}: 0.987
\end{itemize}
Again, we softmax the output and optimize categorical cross entropy. 

\bibliography{inspire}

\end{document}

%% file: arch_diagram.pdf_tex
\begingroup%
  \makeatletter%
  \providecommand\color[2][]{%
    \errmessage{(Inkscape) Color is used for the text in Inkscape, but the package 'color.sty' is not loaded}%
    \renewcommand\color[2][]{}%
  }%
  \providecommand\transparent[1]{%
    \errmessage{(Inkscape) Transparency is used (non-zero) for the text in Inkscape, but the package 'transparent.sty' is not loaded}%
    \renewcommand\transparent[1]{}%
  }%
  \providecommand\rotatebox[2]{#2}%
  \newcommand*\fsize{\dimexpr\f@size pt\relax}%
  \newcommand*\lineheight[1]{\fontsize{\fsize}{#1\fsize}\selectfont}%
  \ifx\svgwidth\undefined%
    \setlength{\unitlength}{1021.14207382bp}%
    \ifx\svgscale\undefined%
      \relax%
    \else%
      \setlength{\unitlength}{\unitlength * \real{\svgscale}}%
    \fi%
  \else%
    \setlength{\unitlength}{\svgwidth}%
  \fi%
  \global\let\svgwidth\undefined%
  \global\let\svgscale\undefined%
  \makeatother%
  \begin{picture}(1,0.26719223)%
    \lineheight{1}%
    \setlength\tabcolsep{0pt}%
    \put(0.49880078,0.2588835){\makebox(0,0)[t]{\lineheight{1.25}\smash{\begin{tabular}[t]{c}\textbf{Particlewise} \eqref{eq:inv_deep_set}\\$ F\left(\Phi(\mathcal X) \right)$ \\\end{tabular}}}}%
    \put(0.16331959,0.25888353){\makebox(0,0)[t]{\lineheight{1.25}\smash{\begin{tabular}[t]{c}\textbf{Nested Concatenation} \eqref{eq:nested}\\$ F^{(L)} \left(\Phi_\oplus^{(L)}\left(\mathcal X\right)  \right)$ \\\end{tabular}}}}%
    \put(0.49880794,0.14867033){\makebox(0,0)[t]{\lineheight{1.25}\smash{\begin{tabular}[t]{c}\textbf{Pairwise} \eqref{eq:pairwise_arch}\\$ F\left(\Phi_2(\mathcal X, \mathcal X) \right)$\end{tabular}}}}%
    \put(0.85763453,0.14867035){\makebox(0,0)[t]{\lineheight{1.25}\smash{\begin{tabular}[t]{c}\textbf{Nonlinear Pairwise} \eqref{eq:edgeconvnl}\\$ F \left(\Phi_2^\Pi(\mathcal X)  \right) $\end{tabular}}}}%
    \put(0.50018054,0.03130149){\makebox(0,0)[t]{\lineheight{1.25}\smash{\begin{tabular}[t]{c}\textbf{Tripletwise} \eqref{eq:triplesum}\\$ F \left(\Phi_3(\mathcal X, \mathcal X,\mathcal X)\right)$\end{tabular}}}}%
    \put(0.85762737,0.03130149){\makebox(0,0)[t]{\lineheight{1.25}\smash{\begin{tabular}[t]{c}\textbf{Iterated Nonlinear Pairwise} \eqref{eq:iteratedecnl}\\$F \left(\Phi_2^{(L),\Pi}\left(\mathcal X^{(L-1)}\right)\right)$\end{tabular}}}}%
    \put(0,0){\includegraphics[width=\unitlength,page=1]{arch_diagram.pdf}}%
    \put(0.49128061,0.07763644){\makebox(0,0)[t]{\lineheight{1.25}\smash{\begin{tabular}[t]{c}$\scriptstyle \Phi_3(\mathcal X, \mathcal X, \mathcal X)=\Phi_2(\mathcal X, \mathcal X)$\end{tabular}}}}%
    \put(0,0){\includegraphics[width=\unitlength,page=2]{arch_diagram.pdf}}%
    \put(0.85539241,0.07733847){\makebox(0,0)[t]{\lineheight{1.25}\smash{\begin{tabular}[t]{c}$\scriptstyle L=1$\end{tabular}}}}%
    \put(0,0){\includegraphics[width=\unitlength,page=3]{arch_diagram.pdf}}%
    \put(0.49105777,0.19078279){\makebox(0,0)[t]{\lineheight{1.25}\smash{\begin{tabular}[t]{c}$\scriptstyle \Phi_2(\mathcal X, \mathcal X)=\Phi(\mathcal X) $\end{tabular}}}}%
    \put(0.37428818,0.250987){\makebox(0,0)[rt]{\lineheight{1.25}\smash{\begin{tabular}[t]{r}$\scriptstyle L=0 $\end{tabular}}}}%
    \put(0,0){\includegraphics[width=\unitlength,page=4]{arch_diagram.pdf}}%
    \put(0.16525313,0.14867035){\makebox(0,0)[t]{\lineheight{1.25}\smash{\begin{tabular}[t]{c}\textbf{Nested Concatenation w/ Memory} \eqref{eq:gen_nested}\\$ F^{(L)} \left(\Phi_\oplus^{(L)}\left(\mathcal X^{(L-1)}\right)  \right)$ \\\end{tabular}}}}%
    \put(0,0){\includegraphics[width=\unitlength,page=5]{arch_diagram.pdf}}%
    \put(0.63501454,0.13938509){\makebox(0,0)[lt]{\lineheight{1.25}\smash{\begin{tabular}[t]{l}$\scriptstyle \Pi(x)=x $\end{tabular}}}}%
    \put(0,0){\includegraphics[width=\unitlength,page=6]{arch_diagram.pdf}}%
  \end{picture}%
\endgroup%